\documentclass[aps,prd,onecolumn,showpacs,nofootinbib,amsmath,amssymb,amsfonts,showkeys]{revtex4}

\usepackage{graphicx}
\usepackage{bm}

\begin{document}

\title{Dynamics of minimally coupled dark energy in spherical halos of dark matter}

\author{Bohdan Novosyadlyj, Maksym Tsizh, Yurij Kulinich}
 
\affiliation{Astronomical observatory of Ivan Franko National University of Lviv,\\ 
              Kyryla i Methodia str., 8, Lviv, 79005, Ukraine\\
              \email{bnovos@gmail.com, tsizh@astro.franko.lviv.ua, kul@astro.franko.lviv.ua}}


\begin{abstract}
 We analyse the evolution of scalar field dark energy in the spherical halos of dark matter at the late stages of formation of gravitationally bound systems in the expanding Universe. The dynamics of quintessential dark energy at the center of dark matter halo strongly depends on
 the value of effective sound speed $c_s$ (in units of speed of light). If $c_s\sim1$ (classical scalar field) then the dark energy in the gravitationally bound systems is only slightly perturbed and its density is practically the same as in cosmological background. The dark energy with small value of sound speed ($c_s<0.1$), on the contrary, is important dynamical component of halo at all stages of their evolution: linear, non-linear, turnaround, collapse, virialization and later up to current epoch. These properties of dark energy can be used for constraining the value of effective sound speed $c_s$ by comparison the theoretical predictions with observational data related to the large scale gravitationally bound systems. 
 \end{abstract}
\pacs{95.36.+x, 98.80.-k}
\keywords{Dark energy -- Gravitational instability -- Large scale structure formation -- Gravitationally bound systems}
\maketitle 

\section{Introduction}
\label{intro}
 The progress of observational cosmology brings us more and more evidence of existence of dark energy, the nature of which is still unknown  
despite of significant efforts of researchers. On the other hand, the  extensive theoretical modelling \cite{yoo,copeland,tsuji,caldwell,Amendola2010,Wolschin2010,Ruiz2010,Novosyadlyj2013m,GRG2008,Amendola2013} of this component of our Universe leaves less and less hope that it is of the simplest type like cosmological constant. This encourages us to a deep analysis of the properties of different types of dark energy and its possible observational manifestations at different space-time scales.  

The dynamics of minimally coupled scalar field dark energy of different types in the expanding Universe is extensively studied (see textbooks  and overviews \cite{Amendola2010,Wolschin2010,Ruiz2010,Novosyadlyj2013m,GRG2008,Amendola2013} and citing therein), its influence on the measurable relations such as the luminosity distance-redshift, the angular distance-redshift or characteristics of the large scale structure of the Universe is reliably computed, which gives the possibility to constrain the values of parameters of specific models of dark energy.  
It is generally accepted, that at cosmological scales the viable models of dark energy are whether strongly homogeneous or slightly perturbed, they are well described in the framework of linear theory of cosmological perturbations. The current data of observational cosmology give the possibility to determine the dark energy density parameter $\Omega_{de}$ with accuracy $\sim2\%$ and equation of state parameter $w_0$ with accuracy of $\sim6\%$. Moreover, they prefer phantom dark energy at $\sim2\sigma$ confidential level \cite{Xia2013,Rest2014,Cheng2014,Shafer2014,Novosyadlyj2014}. However, the other parameter - the effective sound speed $c_s$, which is important also for establishing the nature of dark energy, is not constrained by current data of observational cosmology 
\cite{Sergijenko2015}. So, other sensitive tests should be found.

The dynamics of minimally coupled scalar field dark energy at scales of gravitational bound systems is not studied so extensively as at cosmological scales mainly due to complexity of non-linear dynamics of dark matter and baryon components that needs the numerical N-body-hydro-thermodynamics simulations. The main features and characteristic times of dynamics of dark matter at the late stages of formation of gravitational bound systems are obtained from the analysis of spherical model which is generalized for $\Lambda$CDM-model and dark matter plus unclustered quintessence on scales much smaller than the horizon for different aspects of large scale structure formation \cite{Gunn1972,Press1974,Peebles1980,Bond1991,Bower1991,Lahav1991,Lacey1993,Eke1996,Wang1998,Cooray2002,Weller2002,Battye2003,Kulinich2003,Weinberg2003,Shaw2008,Kulinich2013}. 
In the numerical N-body simulations of large scale structure formation (see overviews \cite{Kuhlen2012,Baldi2012} and citing therein) the dark energy is implemented also mainly for the computations of the correct Hubble function $H(z)$ and initial power spectrum of matter density perturbations $P_i(k)\equiv\langle \delta_m(k)\delta_m^*(k)\rangle$ for the specific model under investigation. The effects of influence of dark energy perturbations on the evolution of matter components at the non-linear stage are weak and usually are not taken into account. 

For clustered quintessence on subhorizon scales the first results for non-linear collapse have been obtained by D. Mota and C. van de Bruck \cite{Mota2004}. They have shown that although the scalar field dark energy is subdominant at the time of virialization of dark matter halo, it significantly alters the predictions of the spherical collapse model for density contrast of matter and its virial radius if the field collapses together with the dark matter. The inclusion of dark energy into the formalism of spherical collapse and virialization of dark matter-dark energy system is analysed in \cite{Maor2005}. Authors of \cite{Manera2006,Nunes2006} have shown that highly perturbed dark energy models present distinct features which may be used to confront them with observational data on number counts of galaxies and clusters. Creminalli et al. \cite{Creminelli2010} studied the spherical collapse model in the presence of quintessence with negligible speed of sound (other
motivations for such dark energy models see in \cite{Lim2010}) which follows dark matter during the collapse and analysed 
how such dark energy affects the dark matter mass function. They have shown that the large part of the mass function computed in $\Lambda$CDM model is distinguishable from the function computed in the model with scalar field dark energy. Two-component (dark matter and dark energy) collapse was studied by Q. Wang and Z. Fan using spherical shell approach \cite{wang2012}.

In this paper we analyse the dynamics of minimally coupled scalar field dark energy at all stages of formation of gravitational bound systems
starting from the linear stage in the early Universe. The general relativistic approach is used for all stages and all components. We suppose that dark energy in the dark matter halo and on the cosmological background is of the same nature and origin and its key parameters are the same.  

\section{Spherical inhomogeneities at FRW background}

\subsection{Cosmological background} 

We assume that cosmological background is the spatially flat, homogeneous and isotropic Universe with
Friedmann-Robertson-Walker (FRW) metric of 4-space
\begin{equation}
ds^2=g_{ij}dx^idx^j=dt^2-a^2(t)[dr^2+r^2(d\theta^2+\sin^2\theta d\varphi^2)] ,\label{ds_cb}
\end{equation}
where the Latin indices $i,\,j,\,...$ run from 0 to 3 and the Greek ones are used for the spatial coordinates: $\alpha,\,
\beta,\,...$=1, 2, 3. Here and below we put the speed of light $c=1$, the scale factor we normalize to 1 at current epoch, $a(t_0)=1$.
We also suppose that the Universe is filled with non-relativistic dust-like matter component (cold dark matter and baryons), relativistic ones (thermal electromagnetic radiation and massless neutrino) and minimally coupled dark energy. We are interested here in the stages of evolution of the Universe when all components with enough accuracy can be described in the perfect fluid approximation with energy densities $\bar{\varepsilon}_{m}(t)$, $\bar{\varepsilon}_{\rm r}(t)$ and $\bar{\varepsilon}_{de}(t)$, pressures $\bar{p}_{m}=0$,  
$\bar{p}_{\rm r}=\bar{\varepsilon}_{\rm r}(t)/3$ and $\bar{p}_{de}=w\bar{\varepsilon}_{de}(t)$ and four-velocities
$\bar{u}^{\alpha}_{m}=\bar{u}^{\alpha}_{\rm r}=\bar{u}^{\alpha}_{de}=0$ accordingly. The last equality means that frame is comoving to Hubble flow at any $t=const$ hypersurface.  We suppose also that dark energy is scalar field with constant effective sound speed $c_s$ and constant parameter of equation of state (EoS)  $w$. 
Assuming the gravitational interaction only between components (minimal coupling) and introducing the common used definitions $\bar{\varepsilon}_{cr}^{(0)}\equiv3H_0^2/{8\pi G}=\bar{\varepsilon}_{m}^{(0)}+\bar{\varepsilon}_{\rm r}^{(0)}+\bar{\varepsilon}_{de}^{(0)}$,
$\Omega_m\equiv\bar{\varepsilon}_{m}^{(0)}/\bar{\varepsilon}_{cr}^{(0)}$, $\Omega_{\rm r}\equiv\bar{\varepsilon}_{\rm r}^{(0)}/\bar{\varepsilon}_{cr}^{(0)}$ and $\Omega_{de}\equiv\bar{\varepsilon}_{de}^{(0)}/\bar{\varepsilon}_{cr}^{(0)}$, where ``(0)'' denotes the current values, we obtain from the conservation law and Einstein equations the well known relations for dynamics of cosmological background
\begin{eqnarray}
&&H(a)=H_0\sqrt{\Omega_{\rm r}a^{-4}+\Omega_ma^{-3}+\Omega_{de}a^{-3(1+w)}},\label{H}\\
&&q(a)=\frac{1}{2}\frac{2\Omega_{\rm r}a^{-4}+\Omega_ma^{-3}+(1+3w)\Omega_{de}a^{-3(1+w)}}
{\Omega_{\rm r}a^{-4}+\Omega_ma^{-3}+\Omega_{de}a^{-3(1+w)}},\label{q}\\
&&\bar{\varepsilon}_m(a)=\bar{\varepsilon}_{cr}^{(0)}\Omega_ma^{-3}, \,\, \bar{\varepsilon}_{\rm r}(a)=\bar{\varepsilon}_{cr}^{(0)}\Omega_{\rm r}a^{-4}, \,\, \bar{\varepsilon}_{de}(a)=\bar{\varepsilon}_{cr}^{(0)}\Omega_{de}a^{-3(1+w)}, \label{rho_b}
\end{eqnarray}
which will be used below. 
Here $H\equiv\frac{d\ln{a}}{dt}$ is the rate of expansion of the Universe (Hubble parameter) and $q\equiv-\frac{d^2a}{dt^2}\frac{1}{aH^2}$ is the dimensionless deceleration parameter. The time variable $t$ is uniquely connected with scale factor $a$ by the integral relation $t=\int_0^a da'/a'H(a')$ and we will use $a$-dependences instead $t$-dependences. 

\subsection{Spherical inhomogeneities}

Subject of this study is the scalar mode of cosmological perturbations generated in the early Universe. Let us analyse the evolution of spherical perturbation from linear stage in early radiation dominated epoch, through quasi-linear stage and turnaround point at matter dominated epoch to highly non-linear stage, infall and formation of spherical dark matter halo at dark energy dominated epoch. The local spherical perturbation distorts the FRW metric (\ref{ds_cb}) so that it becomes
\begin{equation}
ds^2=e^{\nu(t,r)}dt^2-a^2(t)e^{\mu(t,r)}[dr^2+r^2(d\theta^2+\sin^2\theta d\varphi^2)],\label{ds_sph}
\end{equation}
where metric functions $\nu(t,r)$ and $\mu(t,r)$ $\rightarrow0$ at cosmological background. At the linear stage, when $\nu(t,r)\sim\mu(t,r)\ll1$
the metric (\ref{ds_sph}) becomes the metric of conformal-Newtonian (longitudinal) gauge \cite{Bardeen1980} in spherical coordinates, since  $e^{\nu(t,r)}\approx1+\nu$, $e^{\mu(t,r)}\approx1+\mu$. The space-time coordinates for small $\mu$, $\nu$ are practically the same as in cosmological background, that is why we call them FRW frame. One can introduce the local frame based on the proper time interval of the observer at the distance $r$ from center, $d\tau(r)$, which is connected with FRW frame interval $dt$ as $d\tau(r)=e^{\nu(r)/2}dt$, and proper distance interval along radial distance, which is $d{R}=e^{\mu(r)/2}dr$. The angle coordinates of local frame are the same as in FRW frame.
In the metric (\ref{ds_sph}) the 4-velocity of any component can be presented through corresponding 3-velocities and is as follows  
\begin{eqnarray}
u^i(t,r) = \left\{\frac{e^{-\nu/2}}{\sqrt{1-v^2}},\,\,\frac{a^{-1}e^{-\mu/2}v}{\sqrt{1-v^2}},\,\,0,\,\,0\right\},
\label{u-v}
\end{eqnarray}
where the radial component of 3-velocity $v$ is defined as the ratio of proper space interval 
to proper time interval at distance $r$ from the center: $v(\tau,r)\equiv dR/d\tau$ (in units of speed of light). The covariant components of the 4-velocity are: $u_i=g_{ij}u^j$. 
The components of the energy-momentum tensor (EMT) $T_i^k = (\varepsilon+p)u_iu^k - \delta_i^kp$ with energy density $\varepsilon(t,r)$, pressure $p(t,r)$ and 4-velocity (\ref{u-v}) are the following:
\begin{eqnarray}
T_0^0 = \frac{\varepsilon+pv^2}{1-v^2},\quad T_0^1 = \frac{\varepsilon+p}{1-v^2}\frac{e^{(\nu-\mu)/2}}{a}v, 
\quad T_1^1 =-\frac{\varepsilon v^2+p}{1-v^2}, \quad  T_2^2 = T_3^3 = - p.
\label{tei}
\end{eqnarray}
These are exact expressions for components of EMT. However, in this paper we investigate the dynamics of collapse of spherical halo at highly non-linear stage which ends with forming of gravitationally bound virialized structures of scales of galaxy - cluster of galaxies\footnote{The excellent overview of the main properties of galaxies and clusters as well as the standard scenario of their formation one can find in the book \cite{Padmanabhan2002}.}. At these scales the maximal values of peculiar velocities are $\sim10^2-10^3$ km/s, so, we can present $(1-v^2)^{-1}\approx1+v^2$ since $v^2\ll1$. We keep the terms with $v^0,\,v^1,\,v^2$ and omit the terms with $v^3,\,v^4$ for all components. The metric functions $\nu(t,r)$ and $\mu(t,r)$ are the same order as Newtonian gravitational potential, which for interested us scales is essentially lower than squared speed of light, it means that $\nu$, $\mu\ll1$. So, the components of EMT (\ref{tei}) can be simplified to
\begin{eqnarray} 
&&T^0_{0\,(m)} = \varepsilon_m(1+v_m^2),\,\, T^1_{0\,(m)} =a^{-1}\varepsilon_m v_m,\,\,
T^1_{1\,(m)} =-\varepsilon_m v_m^2,\,\,  T^2_{2\,(m)} = T^3_{3\,(m)} = 0, \label{tei_m}\\ \nonumber\\ 
&&T^0_{0\,(de)} = \varepsilon_{de}+(\varepsilon_{de}+p_{de})v_{de}^2,\,\, T^1_{0\,(de)}=a^{-1}(\varepsilon_{de}+p_{de})v_{de}, \nonumber\\
&&T^1_{1\,(de)} = -p_{de}-(\varepsilon_{de}+p_{de})v_{de}^2 ,\,\,  T^2_{2\,(de)} = T^3_{3\,(de)} = - p_{de}, \label{tei_de} \\ \nonumber\\
&&T^0_{0\;(\rm r)}= \varepsilon_{\rm r},\,\, T^1_{0\;(\rm r)} =a^{-1}(\varepsilon_{\rm r}+p_{\rm r})v_{\rm r}, \,\,
T^1_{1\;(\rm r)} = T^2_{2\;(\rm r)} = T^3_{3\;(\rm r)} = - p_{\rm r}, \label{tei_r} 
\end{eqnarray}
where all $\varepsilon$, $p$ and $v$ are functions of $t$ and $r$. The terms with $v_m^2$ and $v_{de}^2$ are important at highly non-linear stage of evolution of overdensity when $\varepsilon_m$ or $\varepsilon_{de}$ reach of large values. 
But we have omitted the similar terms for relativistic component since its density is negligible at the late epoch when structures are forming. 
The energy densities and pressures for spherical perturbation in each component in the FRW frame we present as follows
\begin{eqnarray}
&&\varepsilon_m(t,r)=\bar{\varepsilon}_m(t)(1+\delta_m(t,r)), \quad p_m(t,r)=0, \nonumber\\
&&\varepsilon_{de}(t,r)=\bar{\varepsilon}_{de}(t)(1+\delta_{de}(t,r)), \quad p_{de}(t,r)=
w\bar{\varepsilon}_{de}+\delta p_{de}(t,r), \nonumber \\  
&&\varepsilon_{\rm r}(t,r)=\bar{\varepsilon}_{\rm r}(t)(1+\delta_{\rm r}(t,r)), \quad p_{\rm r}(t,r)=
\frac{1}{3}\bar{\varepsilon}_{\rm r}(t)(1+\delta_{\rm r}(t,r)), \nonumber
\end{eqnarray}
where the variation of pressure of dark energy in the non-comoving to dark energy coordinates contains the adiabatic and 
non-adiabatic parts, so 
$$\delta p_{de}(t,r)=c_s^2\bar{\varepsilon}_{de}\delta_{de}(t,r)-
3\bar{\varepsilon}_{de}aH(1+w)(c_s^2-w)\int{v_{de}(t,r)dr}$$
(see for details Appendix~A). The density perturbations of each component are defined as follows: $\delta_N(t,r)\equiv \varepsilon_N(t,r)/\bar{\varepsilon}_N(t)-1$. Such presentation is useful for setting the initial conditions and comparison with well known evolution of linear 
perturbations. 

To obtain the equations, which describe the evolution of spherical inhomogeneous in all components, we use the Einstein and energy-momentum conservation law differential equations. It is comfortable to write and integrate them in the variables $(a,\,r)$ instead of $(t,\,r)$. 

\subsection{Einstein equations}
We deduce the equations of evolution for metric functions $\nu(a,r)$ and $\mu(a,r)$ from Einstein equations
\begin{eqnarray}
&&R^i_j-\frac{1}{2}\delta^i_j R=\kappa \left(T^i_{j\;(m)}+T^i_{j\;(\rm r)}+T^i_{j\;(de)}\right),\nonumber
\end{eqnarray}
In the general case the left hand side of equations is  quasilinear for metric functions. In this paper we are interested of the evolution of spherical overdensities of scale range of galaxy - cluster of galaxies from the early stages when they are superhorizon and linear to the late stages when they are subhorizon virialized gravitationally bound systems, which are highly non-linear in the matter density perturbation, $\delta_m\gg1$, but linear in the gravitational potential\footnote{At the current epoch the galaxies and clusters are gravitationally classical objects.}, $\nu,\,\mu \ll1$. So, we omit the non-linear terms with $\nu^2,\,\mu^2$, $\dot{\nu}^2,\,\dot{\nu}\mu$ and so on. Therefore, we obtain the system of four independent 
equations.
\begin{eqnarray}
&&3aH^2\dot\mu-\frac{1+\mu}{a^2}\left(\mu''+\frac{2}{r}\mu'\right)
-3H^2\nu=\kappa\delta T^0_0, \label{ee00_l}\\ 
&&\frac{H}{a}\left(\dot\mu'-\frac{1}{a}\nu' \right)=\kappa\delta T^1_0, \label{ee01_l}\\
&&a^2H^2\ddot\mu+aH^2\left[(3-q)\dot\mu-\dot\nu\right]-\frac{(\mu'+\nu')}{a^2r}-H^2(1-2q)\nu=\kappa\delta T^1_1,  \label{ee11_l}\\  
&&a^2H^2\ddot\mu+aH^2\left[(3-q)\dot\mu-\dot\nu\right]-\frac{\nu''+\mu''}{2a^2}-\frac{\nu'+\mu'}{2a^2r}
- H^2(1-2q)\nu=\kappa\delta T^2_2=\kappa\delta T^3_3, \label{ee22_l}
\end{eqnarray}
where $\delta T^i_j$ is the sum of perturbed parts of the EMT's of all components, the functions $H(a)$ and $q(a)$ are defined by relations (\ref{H})-(\ref{q}). A dot denotes the partial derivative with respect to scale factor $a$, 
($\dot{\;\;}$)$\equiv \partial/\partial{a}$, and a prime
marks the partial derivative with respect to $r$, ($\,'\,$)$\equiv \partial/\partial{r}$.	

Let us construct the equation, which is difference of last two ones:
\begin{equation}
\frac{r}{2a^2}\left(\frac{\nu'+\mu'}{r}\right)'=\kappa\left(T^1_1-T^2_2\right).\label{nu-mu}
\end{equation} 
At the linear stage the right hand side of equation equals zero and integration of this equation gives the relation:
\begin{equation}
 \nu(a,r)+\mu(a,r)=r^2f_1(a)+f_2(a), \nonumber
\end{equation} 
where $f_1(a)$ and $f_2(a)$ are arbitrary functions of time (scale factor), which we can put equal zero. So, at the linear stages $\mu=-\nu$.
At the non-linear stage, for which we keep the term $\sim v_m^2$ in (\ref{tei_m}), the right hand side of equation (\ref{nu-mu}) becomes
$\kappa\left(T^1_1-T^2_2\right)=-\kappa\bar{\varepsilon}_m(a)(1+\delta_m(t,r))v_m^2$ which goes to 0 like $r^2$ for overdensities with smooth top, where $v_m\sim r$. So, at the center of spherical overdensity we have again $\mu=-\nu$. In this paper we will analyse the dynamics of dark energy in the central part of spherical overdensity, therefore we assume {$\mu=-\nu$ for whole time. This gives us the possibility to use one equation from (\ref{ee00_l})-(\ref{ee22_l}) for determination of one metric function $\nu(a,r)$. We use the first equation which finally is as follows
\begin{eqnarray}
\frac{1-\nu}{3a^2}\left(\nu''+\frac{2}{r}\nu'\right)-H^2(a\dot\nu+\nu)=
\frac{H^2_0}{a^3}\left(\Omega_m\delta_m+
\Omega_{\rm r}a^{-1}\delta_{\rm r}+\Omega_{de}a^{-3w}\delta_{de}\right). \label{ee00_l2}
\end{eqnarray}
We have kept in (\ref{ee00_l})-(\ref{ee00_l2}) the non-linear term $\sim\nu\Delta\nu$ since it contributes $\sim10\%$ to value of gravitational potential at the central part of very massive virialized halo. Without this term the equation (\ref{ee00_l2}) becomes the classical Poisson equation for the static Universe, for which $a=1$, $H=0$, $3H_0^2=8\pi G\bar{\rho}_{cr}$ (reverse change of variables), which is good approximation at scales of galaxy - cluster of galaxies when they are deeply subhorizon objects.

\subsection{Equations of the energy-momentum conservation}

The \textit{energy conservation law} $T^k_{0\,;k}=0$ (continuity equation) for dark energy in the metric (\ref{ds_sph}) with additional condition $\mu=-\nu$ is as follows
\begin{eqnarray}
&&\dot{\delta}_{de}+\frac{3}{a}(c_s^2-w)\delta_{de}+(1+w)\left[\frac{v'_{de}}{a^2H}+\frac{2v_{de}}{a^2Hr}
-9H(c_s^2-w)\int{v_{de}dr}-\frac{3}{2}\dot{\nu}\right] \nonumber \\
&&\hskip2cm+(1+c_s^2)\left[\frac{\delta'_{de}v_{de}}{a^2H} +\frac{\delta_{de}}{a^2H}\left(v'_{de}+
\frac{2}{r}v_{de}\right)-\frac{3}{2}\delta_{de}\dot{\nu}\right]=0.
\label{de_cl0}
\end{eqnarray}
The terms with $v^2$, $\mu v$,  $\nu v$ and so on are omitted, since they are the values of the second and higher order of smallness in comparison with those which are kept. The continuity equation for dark matter can be obtained from (\ref{de_cl0}) by putting $c_s^2=w=0$ and for relativistic component by putting $c_s^2=w=1/3$
 \begin{eqnarray}
&&\dot{\delta}_{m}-\frac{3}{2}(1+\delta_{m})\dot{\nu}+\frac{1+\delta_{m}}{a^2H}\left(v'_{m}+\frac{2}{r}v_{m}\right)+  
 \frac{\delta'_{m}v_{m}}{a^2H}=0,\label{m_cl0}\\
&&\dot{\delta}_{\rm r}-2(1+\delta_{\rm r})\dot{\nu}+\frac{4}{3}\frac{1+\delta_{\rm r}}{a^2H}\left(v'_{\rm r}+\frac{2}{r}v_{\rm r}\right)+  
 \frac{4}{3}\frac{\delta'_{\rm r}v_{\rm r}}{a^2H}=0.\label{rel_cl0}
\end{eqnarray}

The \textit{momentum conservation law} $T^k_{1\,;k}=0$ for dark energy in the metric (\ref{ds_sph}) gives the equation of motion  
\begin{eqnarray}
&&\dot{v}_{de}+(1-3c_s^2)\frac{v_{de}}{a}+\frac{c_s^2\delta'_{de}}{a^2H(1+w)}+ 
\left(1+\frac{1+c_s^2}{1+w}\delta_{de}\right)\frac{2v_{de}}{a^2H} \left(v_{de}'+\frac{v_{de}}{r}\right)+\nonumber\\
&&\frac{\nu'}{2a^2H}+\frac{1+c_s^2}{1+w}\left[\dot{\delta}_{de}v_{de}+\delta_{de}\dot{v}_{de}+(1-3w)\frac{\delta_{de}}{a}v_{de}+\frac{\nu'\delta_{de}}{2a^2H}\right]=0.
\label{de_cl1}
\end{eqnarray}
The equations of motion for dark matter and for relativistic component can be obtained from (\ref{de_cl1}) by putting $c_s^2=w=0$ and $c_s^2=w=1/3$ accordingly
\begin{eqnarray}
&&\dot{v}_{m}+\frac{v_{m}}{a}+\frac{\nu'}{2a^2H}+\frac{2v_m}{a^2H}\left(v'_m+\frac{v_m}{r}\right)+
\frac{\dot{\delta}_{m}v_{m}}{1+\delta_{m}}=0,\label{m_cl1}\\
&&\dot{v}_{\rm r}+\frac{\nu'}{2a^2H}+\frac{\delta'_{\rm r}}{4a^2H(1+\delta_{\rm r})}+
\frac{\dot{\delta}_{\rm r}v_{\rm r}}{1+\delta_{\rm r}}=0.
\label{rel_cl1}
\end{eqnarray}
In the last equation we have omitted the term similar to forth one in (\ref{m_cl1}), since for relativistic component it is negligible 
always\footnote{The density and velocity perturbations of thermal radiation and neutrino start to decay due to Silk and collisionless damping on the interesting scales soon after entering the particle horizon \cite{Silk1968,Bond1983}}. Therefore, we have 
the system of seven partial differential equations (\ref{ee00_l2})-(\ref{rel_cl1}) for seven unknown functions $\nu(a,r)$, 
$\delta_m(a,r)$, $v_m(a,r)$,  $\delta_{de}(a,r)$, $v_{de}(a,r)$,  $\delta_{\rm r}(a,r)$, $v_{\rm r}(a,r)$ which we solve numerically 
for given initial conditions. 

Basing on our numerical experiments we assure that omitted the non-linear terms\footnote{The equations with all non-linear terms are presented in \cite{Tsizh2015}.} in the equations (\ref{ee00_l2})-(\ref{rel_cl1}) change the results no more than $\sim0.5\%$ at the highest non-linear stages of evolution of spherical halo which we analyse here.

\subsection{Initial conditions}
We analyse the evolution of cosmological perturbations generated in the early Universe.
At the early epoch the amplitudes of cosmological perturbations of space-time metric, densities and velocities of all components are small and the system of partial differential equations (\ref{ee00_l2})-(\ref{rel_cl1}), which describes their evolution, can be linearized for all unknown functions. 
We suppose that profiles of cosmological perturbations are spherical-symmetric in all components and are the same when they are superhorizon.
The profile function can be expanded into series of some orthogonal functions, e.g. spherical ones in our case, which are eigenfunctions of the Laplacian \cite{Durrer2008}. In particular, we can present the perturbations $\nu(a,r)$, $\delta(a,r)$ and $v(a,r)$ by Fourier-Bessel decomposition as infinite sum of Bessel functions $j_{\ell}(kr)$ integrated in Fourier space with corresponding amplitudes $\nu_{\ell\,k}(a)$, $\delta_{\ell\,k}(a)$ and $v_{\ell\,k}(a)$ 
\cite{Leistedt2011}. Here and below $k$ is wave number in the coordinates which are comoving to the cosmological background. Substituting such decomposition into the linearized equations (\ref{de_cl1})-(\ref{rel_cl1}) and collecting the terms near the same $k$ one can obtain the equations for amplitudes of each $\ell,\,k$-mode for each component. For simplicity we will analyse here the equations for the main harmonics $j_0$ for density perturbations which have smooth top in the central point and the first non-zero term 
in the decomposition of velocity perturbations\footnote{In the decomposition of velocity we put $v_{0\, k}(a)=0$ over spherical symmetry: the central point is immovable and near the central point $v(a,r)\propto r$.} with $j_1(kr)=-j_0'(kr)/k$. Therefore, we analyse the perturbations of the metric, density and velocity of any component as follows
\begin{eqnarray}
&&\nu(a,r)=\tilde{\nu}(a)f_k(r)=\tilde{\nu}(a)\frac{\sin{kr}}{kr}, \quad \delta_{N}(a,r)=\tilde{\delta}_{N}f_k(r)=\tilde{\delta}_{N}(a)\frac{\sin{kr}}{kr}, \nonumber \\
&&v_{N}(a,r)=\tilde{v}_{N}(a)f_k'(r)=\tilde{v}_{N}(a)k\left(\frac{\cos{kr}}{kr}-\frac{\sin{kr}}{k^2r^2}\right). \label{profile}
\end{eqnarray}
Here and below the $\ell,\,k$-indexes of amplitudes are substituted by the tilde.} The system of linear ordinary differential equations for amplitudes $\tilde{\nu}(a)$, $\tilde{\delta}_m(a)$, $\tilde{v}_m(a)$, $\tilde{\delta}_{de}(a)$, $\tilde{v}_{de}(a)$, 
$\tilde{\delta}_{\rm r}(a)$, $\tilde{v}_{\rm r}(a)$ are presented in the Appendix~B. We will use them to find the relations between amplitudes at some $a_{init}\ll1$ when the scales of interested here gravitationally bound systems were essentially larger than horizon scale, $a_{init}k^{-1}\gg t$. Obviously, that for galaxy - cluster scales it was early radiation-dominated epoch, when $\bar{\varepsilon}_{\rm r}\gg\bar{\varepsilon}_{m}\gg\bar{\varepsilon}_{de}$, so, the matter and dark energy can be treated as test components. The amplitude of metric function $\tilde{\nu}$ is defined by density perturbations of the relativistic component (see eq. (\ref{ee00_l2l}) in Appendix~B). The non-singular solution of eqs. (\ref{rel_cl0l}) and (\ref{ee00_l2l}) has asymptotic values at $a_{init}$
\begin{equation}
\tilde{\nu}^{init}=-C_k, \quad \tilde{\delta}^{init}_{\rm r}=C_k, \quad \tilde{v}^{init}_{\rm r}=\frac{C_k}{4a_{init}H(a_{init})}, \label{dr_ini}  
\end{equation}
where $C_k$ is some constant (see for details \cite{Novosyadlyj2007}). The solutions of equations for matter (\ref{m_cl0l}) and dark energy (\ref{de_cl0l})-(\ref{de_cl1l}) as test components give the asymptotic values for superhorizon perturbations at $a_{init}$
\begin{eqnarray}
&&\tilde{\delta}_{m}^{init}=\frac{3}{4}C_k,  \quad \quad \quad \quad \tilde{v}_{m}^{init}=\frac{C_k}{4a_{init}H(a_{init})} ,  \label{dm_ini}\\
&&\tilde{\delta}_{de}^{init}=\frac{3}{4}(1+w)C_k,  \quad \tilde{v}_{de}^{init}=\frac{C_k}{4a_{init}H(a_{init})}. \label{de_ini}
\end{eqnarray}
The relations (\ref{dr_ini})-(\ref{de_ini}) are the same for any $\ell,\,k$-mode and have a constants $C_k$ which value specify the initial amplitudes of perturbations for all components. 

\section{Dynamics of dark energy at the center of collapsing halo}

The straightforward way of analysing the evolution of spherical perturbations in three-component medium at non-linear stages consists in integration of system of partial differential equations (\ref{ee00_l2})-(\ref{rel_cl1}). However, in this paper we concentrate our attention on the dynamics of dark energy in the central part of perturbations, which is approximately homogeneous. In the vicinity of center of halo with density profile $f_k(r)=\sin{kr}/kr$ we can substitute the $r$-dependences of unknown function  by terms of their Taylor series of zero and first powers of $r$, so, that 
\begin{eqnarray}
&&f_k(r)\approx1, \quad f_k'(r)\approx-\frac{1}{3}k^2r,  \quad f_k''(r)+\frac{2}{r}f_k'(r)=-k^2f_k(r)\approx-k^2, \nonumber \\
&&F_k(r)\equiv f_k''(r)+\frac{1}{r}f_k'(r)=-k^2f_k(r)-\frac{1}{r}f_k'(r)\approx-\frac{2}{3}k^2. \label{profile2}
\end{eqnarray}
Since the density profile (\ref{profile}) is a smooth function near the center we suppose that space dependences (\ref{profile2}) remain the same on the stage of collapse before virialization. It gives us the possibility to transform the system of partial differential equations (\ref{ee00_l2})-(\ref{rel_cl1}) into the system of ordinary differential equations at $r=0$ 
\begin{eqnarray}
&&\dot{\tilde{\delta}}_{m}-\frac{3}{2}(1+\tilde{\delta}_{m})\dot{\tilde{\nu}}-\frac{1+\tilde{\delta}_{m}}{a^2H}k^2\tilde{v}_m=0,\label{m_cl0o}\\
&&\dot{\tilde{v}}_{m}+\frac{\tilde{v}_{m}}{a}+\frac{\tilde{\nu}}{2a^2H}+
\frac{\dot{\tilde{\delta}}_{m}\tilde{v}_{m}}{1+\tilde{\delta}_{m}}-\frac{4k^2\tilde{v}^2_m}{3a^2H}=0,\label{m_cl1o}\\
&&\dot{\tilde{\delta}}_{de}+\frac{3}{a}(c_s^2-w)\tilde{\delta}_{de}-(1+w)\left[\frac{k^2\tilde{v}_{de}}{a^2H}
+9H(c_s^2-w)\tilde{v}_{de}+\frac{3}{2}\dot{\tilde{\nu}}\right] \nonumber \\
&&\hskip2cm-(1+c_s^2)\left[\frac{k^2\tilde{\delta}_{de}\tilde{v}_{de}}{a^2H}
+\frac{3}{2}\tilde{\delta}_{de}\dot{\tilde{\nu}}\right]=0, \label{de_cl0o}\\
&&\dot{\tilde{v}}_{de}+(1-3c_s^2)\frac{\tilde{v}_{de}}{a}+\frac{c_s^2\tilde{\delta}_{de}}{a^2H(1+w)}+
\frac{\tilde{\nu}}{2a^2H}-\frac{4k^2\tilde{v}^2_{de}}{3a^2H}+\nonumber\\
&&\hskip1cm \frac{1+c_s^2}{1+w}\left[\dot{\tilde{\delta}}_{de}\tilde{v}_{de}+\tilde{\delta}_{de}\dot{\tilde{v}}_{de}+
(1-3w)\frac{\tilde{\delta}_{de}}{a}\tilde{v}_{de}+\frac{\tilde{\nu}\tilde{\delta}_{de}}{2a^2H}\right]=0, \label{de_cl1o}\\
&&\dot{\tilde{\delta}}_{\rm r}-2(1+\tilde{\delta}_{\rm r})\dot{\tilde{\nu}}-\frac{4}{3}\frac{1+\tilde{\delta}_{\rm r}}{a^2H}k^2\tilde{v}_{\rm r}=0,\label{rel_cl0o}\\
&&\dot{\tilde{v}}_{\rm r}+\frac{\tilde{\nu}}{2a^2H}+\frac{\tilde{\delta}_{\rm r}}{4a^2H(1+\tilde{\delta}_{\rm r})}+
\frac{\dot{\tilde{\delta}}_{\rm r}\tilde{v}_{\rm r}}{1+\tilde{\delta}_{\rm r}}=0, \label{rel_cl1o}\\
&&\dot{\tilde{\nu}}+\left(1+(1-\tilde{\nu})\frac{k^2}{3a^2H^2}\right)\frac{\tilde{\nu}}{a} = 
-\frac{\Omega_m\tilde{\delta}_m+\Omega_{\rm r}a^{-1}\tilde{\delta}_{\rm r}+\Omega_{de}a^{-3w}\tilde{\delta}_{de}}
{\Omega_ma+\Omega_{\rm r}+\Omega_{de}a^{1-3w}}. \label{ee00_l2o} 
\end{eqnarray}
For integration of the system
(\ref{m_cl0o})-(\ref{ee00_l2o}) we designed the FORTRAN code dedmhalo.f\footnote{It is available at 
http://194.44.198.6/$\sim$novos/dedmhalo.tar.gz} in which the publicly available subroutine dverk.f \cite{dverk} is used. The results of integration of this system of equations for $r\ll k^{-1}\ne0$ are close to the results for $r=0$, they lead to the same conclusions, therefore, we will not distinguish them here.
\begin{figure*}
\includegraphics[width=0.48\textwidth]{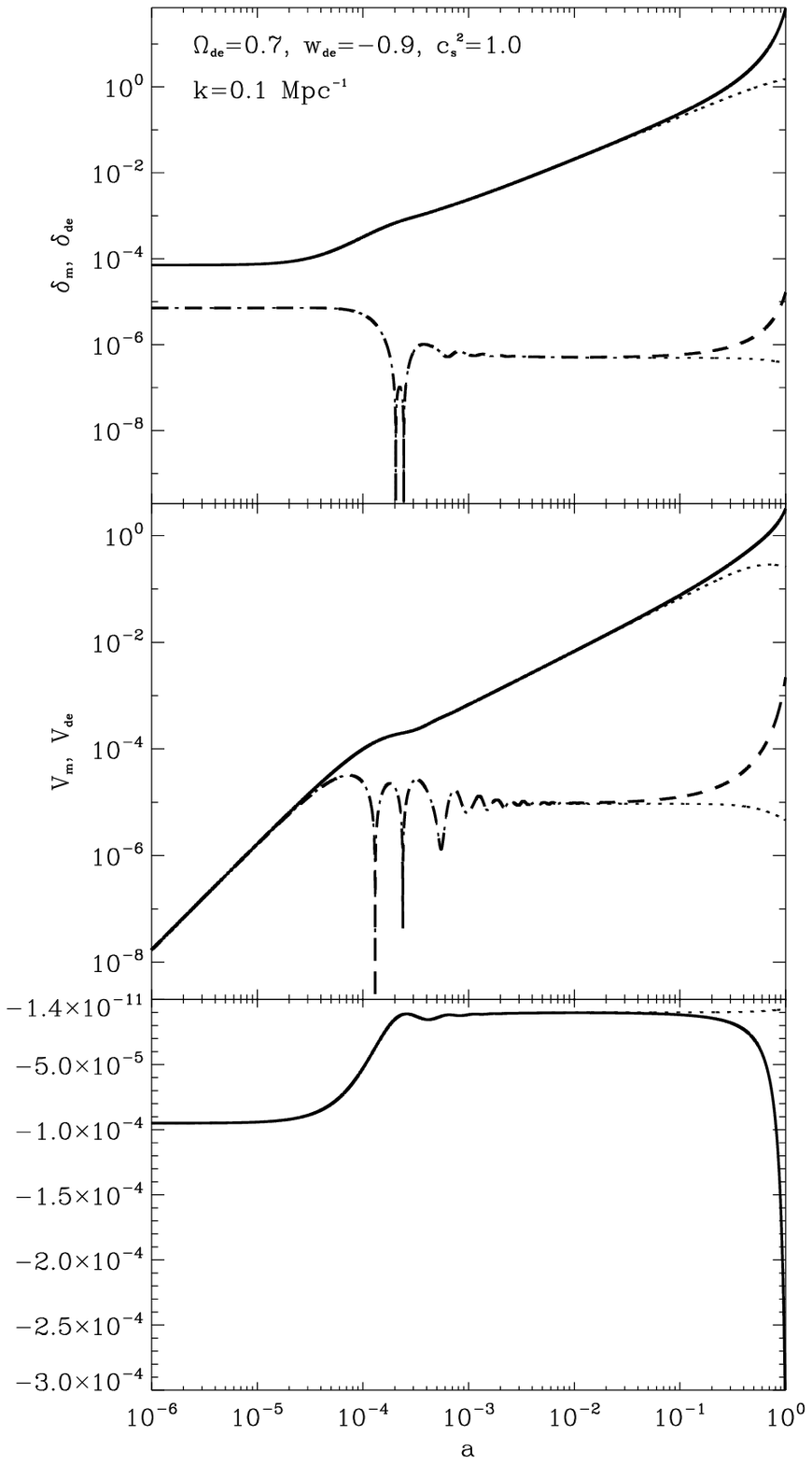}
\includegraphics[width=0.48\textwidth]{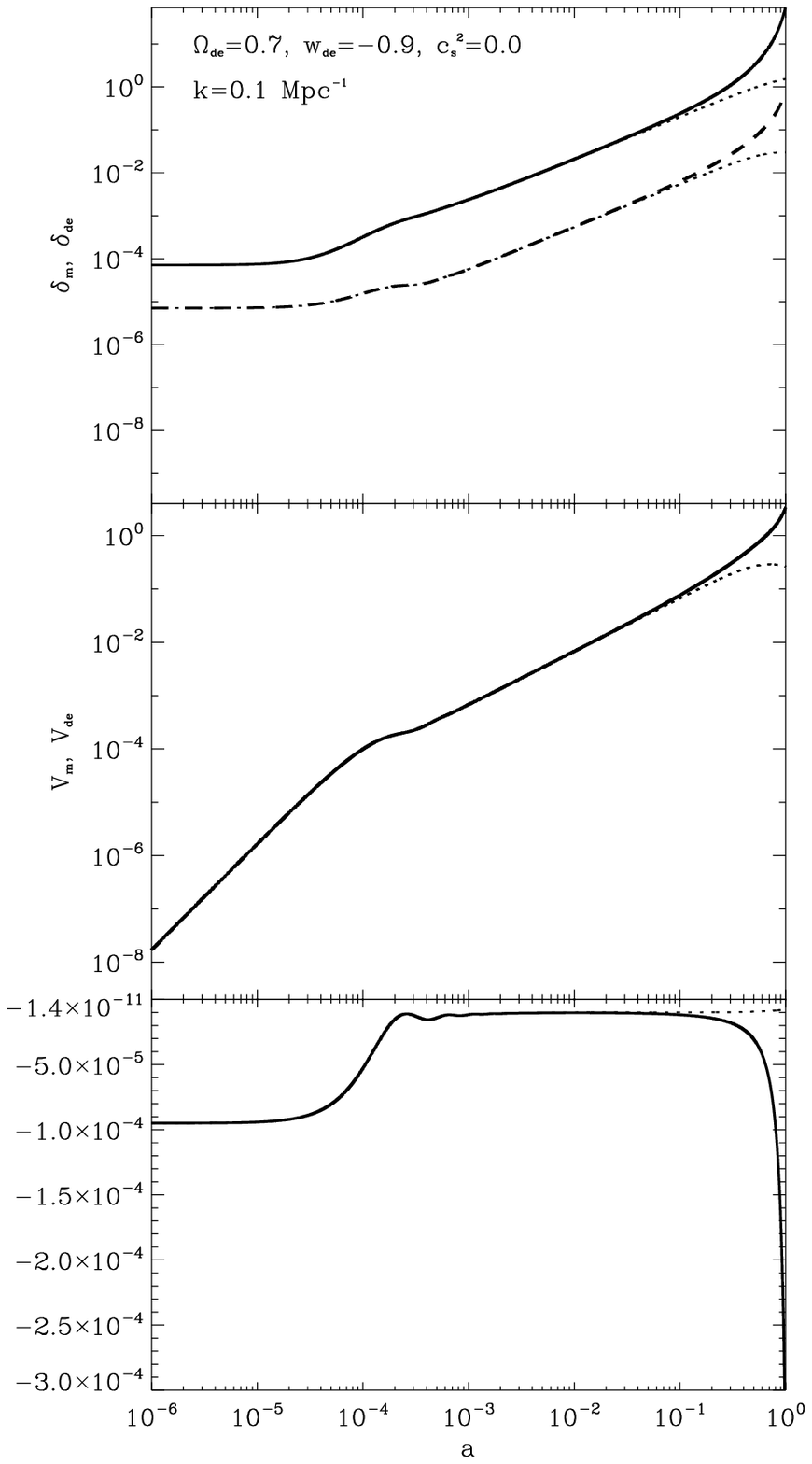}
\caption{Evolution of amplitudes of density perturbations of matter $\tilde{\delta}_m$ and dark energy $\tilde{\delta}_{de}$ (top panels), velocity of matter $V_m$ and dark energy $V_{de}$ in units of Hubble one (middle panels) and gravitational potential $\tilde{\nu}$ (bottom panels) at the central part of spherical halo which is collapsing now. In the top and middle panels solid lines corresponds to matter, dashed lines to dark energy and dotted lines at all panels show the predictions of linear theory.}
\label{collapse}
\end{figure*}

Using this code we calculate the evolution of amplitudes of large-scale spherical perturbation ($k=0.1$ Mpc$^{-1}$, $C_k=9.5\cdot10^{-5}$) which is collapsing at current epoch\footnote{The Great Attractor or alike structures in our Universe are modelled by such perturbations to elucidate their main properties (see for example \cite{Hnatyk1995}).}. The scale $k=0.1$ corresponds to superclusters - largest objects assumed to be gravitationally bounded. From the other hand, objects of smaller scales are already virialized and hence are already not collapsing. We consider such scales in the next section. The results for two models of dark energy with extreme marginal values of effective sound speed 
are presented in Fig. \ref{collapse}. The amplitude of matter density perturbation increases after entering the horizon from initial value $\tilde{\delta}_{m}^{init}=7.125\cdot10^{-5}$ at $a_{init}=10^{-10}$ to $\tilde{\delta}_{m}^{0}=67.9$ at current epoch ($a=1$) in the top left panel and to $\tilde{\delta}_{m}^{0}=72.7$ in the top right panel. The models differ by value of effective sound speed of dark energy only. In the left panel it equals 1, in the right one it equals 0. The evolution of amplitude of density perturbation of dark energy is shown by dashed lines, and one can see, that they differ drastically after entering the horizon: in the case of $c_s^2=1$ it oscillates some time and grows at non-linear stage of  evolution of matter density perturbation,
while in the case of $c_s^2=0$ it increases similar to $\tilde{\delta}_{m}(a)$ and reaches the value $0.83$ at $a=1$. So, in the model of dark energy with vanishing effective sound speed the amplitude of its density perturbation can be high and can affect the collapse of matter at highly non-linear stage. In the presented in Fig. \ref{collapse} case the influence is about $\sim7\%$ only. In the models of dark energy with $c_s^2\sim1$ the amplitude of its density perturbation is negligible small even at stage of collapse of dark matter (top left panel).

The comparison of velocity amplitudes explains the behaviour of density perturbations at the linear and non-linear stages. It is useful to analyse them in units of Hubble velocity, $v_H=aHr$, since in the vicinity of center of halo (see eqs. \ref{profile2}) such ratios do not depend on radial distance\footnote{Since $v_m\ll1$ $v_H\ll1$ we ignore the relativistic transformations of 3-velocities.}:
$$V_m\equiv\frac{v_m}{v_H}\approx-\frac{1}{3}\frac{k^2\tilde{v}_m}{aH}, \quad V_{de}\equiv\frac{v_{de}}{v_H}\approx
-\frac{1}{3}\frac{k^2\tilde{v}_{de}}{aH}.$$
The evolution changes of their absolute values are presented in the middle panels of Fig. \ref{collapse}. For the dark matter it monotonically increases in both cases and reaches the value $3.2$ at current epoch. When $V_m<1$ the perturbation expands, when $V_m>1$ it collapses. The turnaround point of matter, $V_m=1$, is at $a_{ta}\approx0.66$ ($z_{ta}\approx0.52$). The amplitude of matter density perturbation at this time computed by code for integration of non-linear system (\ref{m_cl0o})-(\ref{ee00_l2o}) is $6.784$, while the amplitude calculated by code for linear system (\ref{m_cl0l})-(\ref{ee00_l2l}) equals $1.172$. 
The complete overlapping of solid and dashed lines in the model with $c_s^2=0$ (right middle panel) means that dark energy moves together with dark matter at all stages including turnaround point and collapse. The amplitude of dark energy density perturbation in this model at turnaround point is $\approx0.12$. The ratio of densities of dark energy and matter for such
perturbation at this moment $g_{ta}\equiv\varepsilon_{de}(a_{ta},0)/\varepsilon_{m}(a_{ta},0)$ equals $\approx0.11$. 
For the perturbation with smaller initial amplitude of density perturbation which reaches of turnaround point now ($z_{ta}=0$) this ratio is 
$\approx0.22$. For the perturbation with larger initial amplitude of density perturbation which reaches of turnaround point at $z_{ta}=2$   $g_{ta}\approx0.022$. All three values well agree with $q(z_{ta})$ presented in Fig. 1 in the paper \cite{Maor2005}. 

The evolution of gravitational potentials, shown in the bottom panels, is quite similar in both models. The change of its amplitude in the period between $a=1\cdot10^{-5}$ and $a=5\cdot10^{-4}$ is caused by entering of perturbation into cosmological horizon at the radiation dominated epoch. The acoustic oscillations of relativistic component is noticeable at $a\sim2-5\cdot10^{-4}$. At the matter dominated epoch it is constant and begins fast change at the non-linear stage. We see the formation of gravitational well of spherical halo at the final stage of collapse. We should also note that density perturbations will eventually diverge in such system, since nothing prevents them growing to infinity in finite time.

The results of integration of the linearized system of equations (\ref{m_cl0l})-(\ref{ee00_l2l}) are shown by dotted lines in all panels. 
One can see, that noticeable deviation of non-linear amplitudes and linear ones becomes visible at $a\sim0.1$, long before $a_{ta}$.

\begin{figure*}
\includegraphics[width=0.32\textwidth]{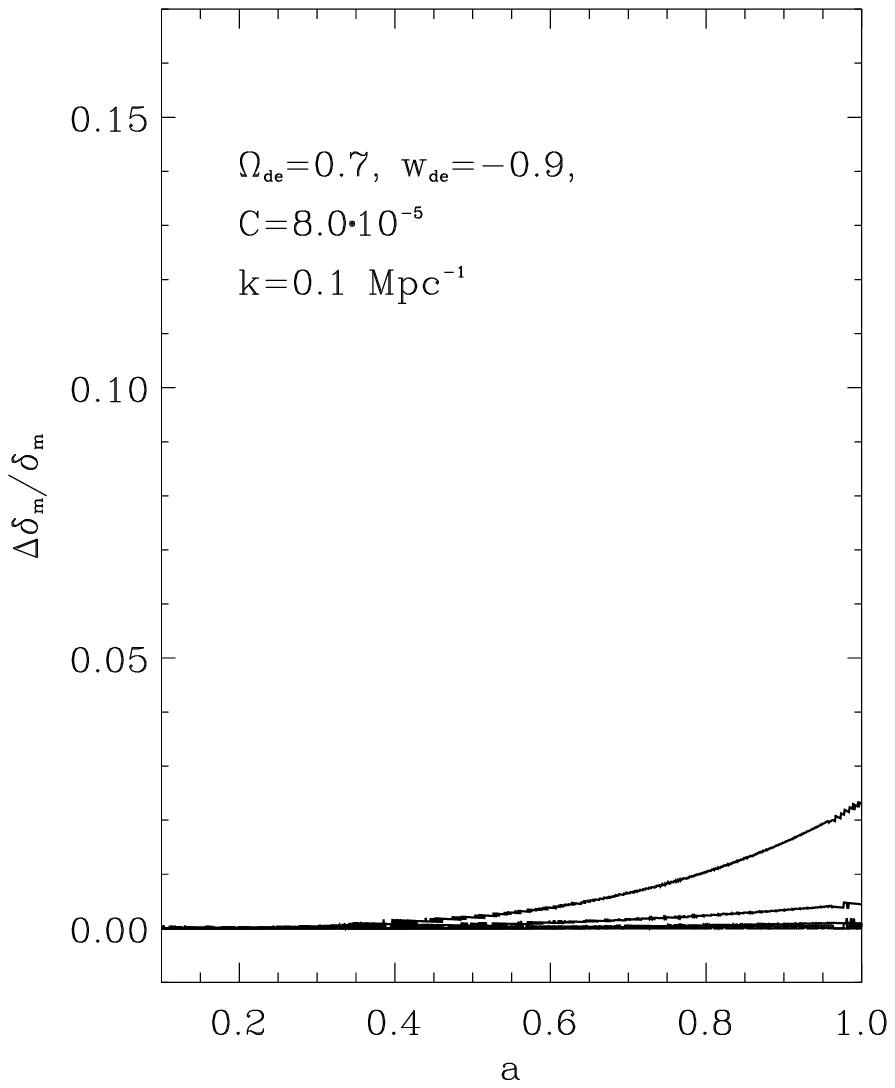}
\includegraphics[width=0.32\textwidth]{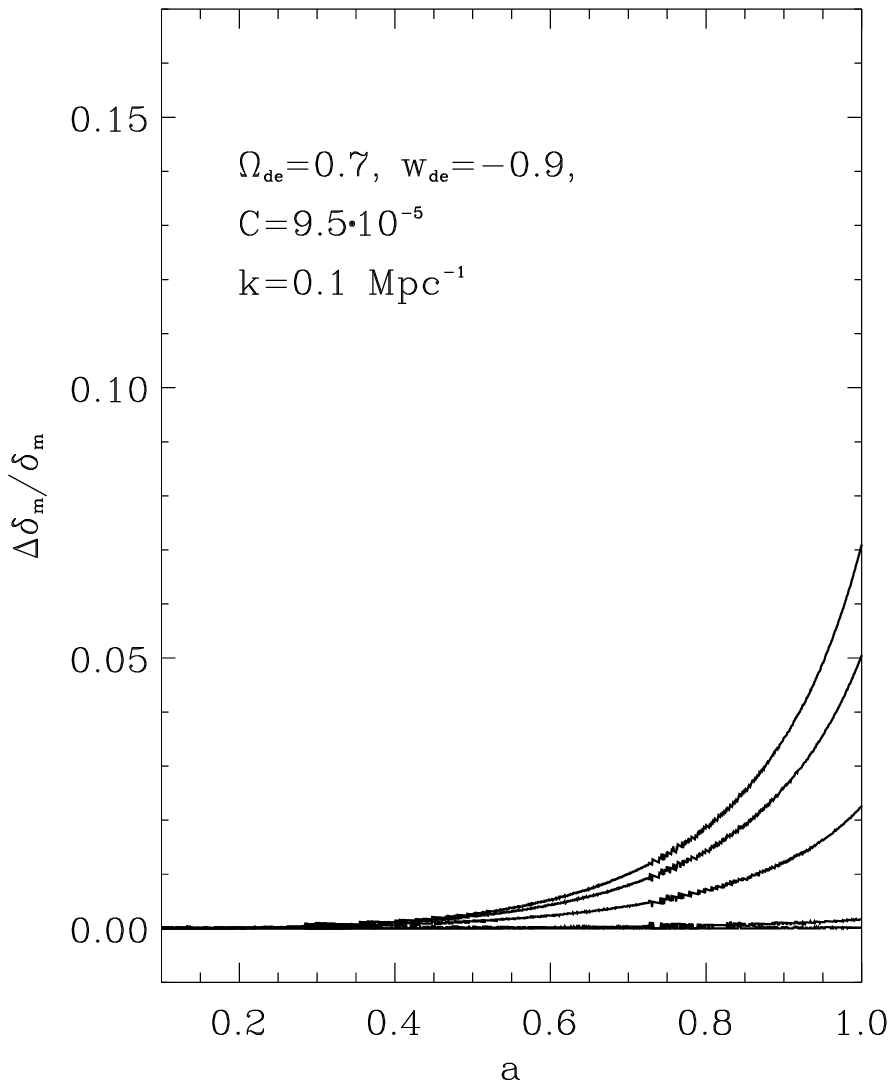}
\includegraphics[width=0.32\textwidth]{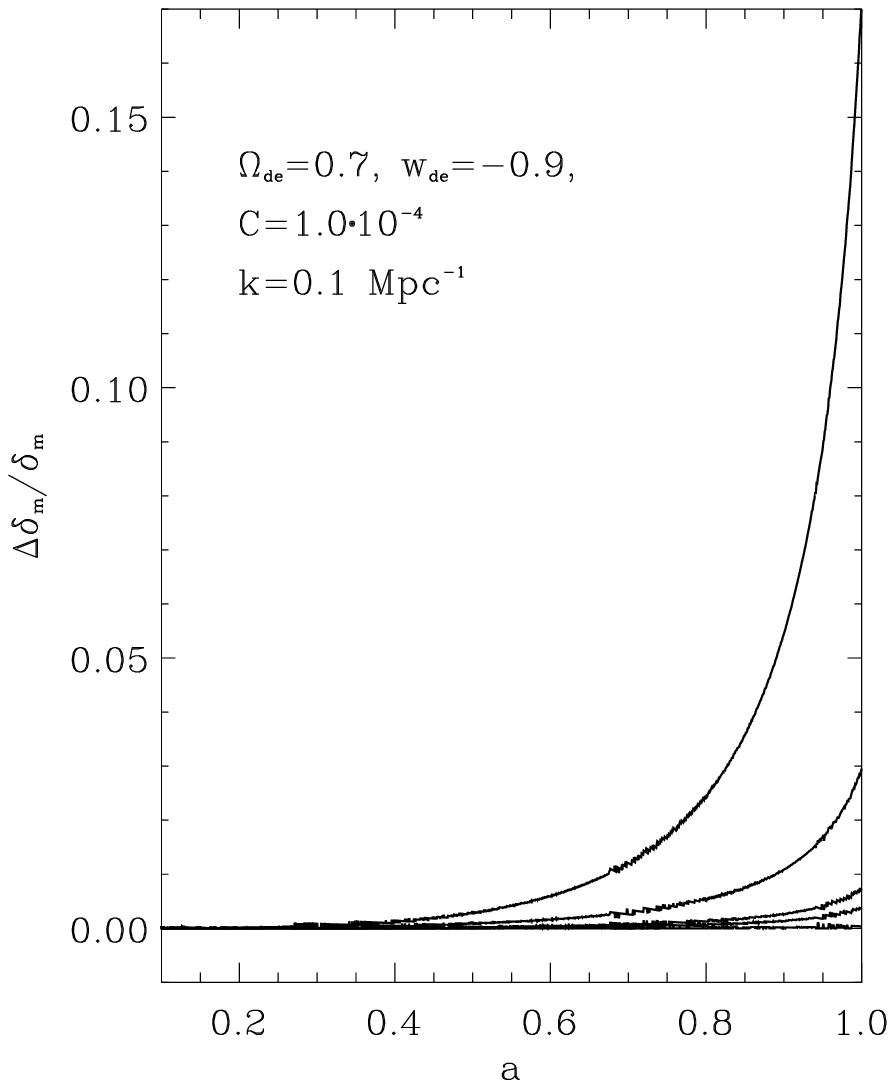}
\caption{The relative differences of matter density perturbations $\Delta\tilde{\delta}_m/\tilde{\delta}_m\equiv  (\tilde{\delta}_m(a;c_s)-\tilde{\delta}_m(a;c_s=1)/\tilde{\delta}_m(a;c_s=1)$ in the cosmological models with different values of effective sound speed ($c_s^2=0.1,\,0.032,\,0.022,\,0.01,\,0.0$ upward in each panel) and different initial amplitudes $C_k$ (panel).}
\label{drho}
\end{figure*}
 
For illustration of gravitation influence of density perturbation of dark energy with different values of $c_s$  on the dark matter one at the non-linear stage of its evolution we present in Fig. \ref{drho} the relative differences of matter density perturbations in the cosmological models with different values of effective sound speed of dark energy 
$\Delta\tilde{\delta}_m/\tilde{\delta}_m\equiv  (\tilde{\delta}_m(a;c_s)-\tilde{\delta}_m(a;c_s=1))/\tilde{\delta}_m(a;c_s=1)$ when all other parameters of cosmological model are the same. The amplitudes of initial perturbations, which are set by $C_k$, are different in the different panels and increase from the left panel to the right one. It means that final amplitudes are different too. In the left panel  at $a=1$ the matter density perturbations in different models of dark energy are near turnaround point: $\tilde{\delta}_m=11.28$, $V_m=1.0$. Decreasing of $c_s$  increases the final amplitudes at $a=1$ and maximal effect is obtained for dark energy model with $c_s=0$: $\tilde{\delta}_m=11.54$, $V_m=1.03$. Therefore, the maximal effect in this case is not larger\footnote{For comparison with linear stage see Fig. 2 in \cite{Sergijenko2015}.} of 3\%. But when the initial amplitudes are larger then effect becomes larger too. In the middle panel the matter in 
the central part of halo is infalling now and has the following values of amplitude of perturbations in the dark energy model with $c_s=1$: $\tilde{\delta}_m=67.9$, $V_m=3.2$.  For lower $c_s$ they are larger and for $c_s=0$ the maximal influence is $\sim6-7\%$: $\tilde{\delta}_m=72.7$, $V_m=3.4$. In the right panel the corresponding values
are as follows $\tilde{\delta}_m=281.2$, $V_m=7.4$ for model with $c_s=1$ and $\tilde{\delta}_m=329.5$, $V_m=8.2$ for model with $c_s=0$.
The influence is $\sim17\%$. Therefore, at the highly non-linear stage of evolution of spherical halo the essential influence of dynamical dark energy on the matter parameters (its density and velocity) is expected when the value of effective sound speed of dark energy is very close to zero. 

Finally, let us show, that considered here values of initial perturbations lead to amplitudes of perturbations in our epoch close observable ones. The r.m.s. value of matter density perturbation at $\sim 8$ ${\rm h^{-1}}$Mpc computed for the initial power spectrum normalised to CMB temperature fluctuations is $\sigma_8=0.815\pm0.009$ \cite{Planck2015a}. Note, that this amplitude is for linear theory and is computed for profile, different from ours (tophat sphere). The amplitudes of matter density perturbation obtained by integration of non-linear system of equations (\ref{de_cl0o})-(\ref{ee00_l2o}) are $\delta^{non-lin}_m=11.28,\,67.9,\,281.2$ for initial conditions with $C_k=8\cdot10^{-5},\,9.5\cdot10^{-5},\,1\cdot10^{-4}$ accordingly. The integration of the linear system of equations (\ref{de_cl0l})-(\ref{ee00_l2l}) for the same initial conditions gives $\delta^{lin}_m=1.28,\,1.52,\,1.60$ correspondingly. After averaging the density perturbation with such linear amplitudes and profile (\ref{profile}) in 
the top-hat sphere of halo radius $R_h=\pi/k$, 
$\bar{\delta}^{lin}_m=3\delta^{lin}_m\int_0^{R_h} f(r)r^2dr/R_h^3$, we obtain: $\bar{\delta}^{lin}_m=0.39,\,0.46,\,0.49$. 
For comparison of these values with measured $\sigma_8$ we should to compute the r.m.s. value of density perturbation at 
$k=0.1$ Mpc$^{-1}$ in the same cosmology. For this we have computed the 
power spectrum of matter density perturbations $P_m(k)$ for given cosmology using code CAMB \cite{camb} and then computed the r.m.s. 
amplitude of density fluctuations at the scale of halo $\sigma_h=\left(\int_0^\infty P_m(k)W^2(kR_h)k^2dk/2\pi^2\right)^{1/2},$
where $W(kR_h)$ is window function of halo of radius $R_h$. For the halo with $k=0.1$ Mpc$^{-1}$ ($R_h=\pi/k_h\approx22$ ${\rm h^{-1}}$Mpc) and top-hat window function we obtain $\sigma_h\approx0.37$. Therefore, the progenitors of spherical halos in Fig. \ref{drho} are peaks of height $\bar{\delta}^{lin}_m/\sigma_h\approx1.1 -- 1.3$ in the random Gaussian field of the initial density fluctuations \cite{Bardeen1986}. Thus, those peaks are widespread and amplitudes of perturbations in structures of considered scales at different stages of non-linear evolution of their central parts are quite usual in the observable Universe.

\section{Dynamics of dark energy at the center of virialized halo}

Now we elucidate the question what is dynamics of dark energy in the gravitationally bound system virialized in the epoch of formation of galaxies and clusters. We set the initial amplitudes for spherical perturbation (\ref{profile}) with $k=1$ Mpc$^{-1}$ in the model with the same cosmological and dark energy parameters as in previous section. At the central part of such perturbation the matter reaches the point of turnaround ($V_m=1$) at $a_{ta}=0.052$ ($z_{ta}\approx18$) when the amplitude of density perturbation equals\footnote{The linear theory (\ref{m_cl0l})-(\ref{ee00_l2l}) for the same time gives 1.064.} 4.58. The amplitude of dark energy density perturbation at $a_{ta}$ depends on $c_s$: for $c_s=1$ it equals $2.2\cdot10^{-7}$, for $c_s=0$ it equals $0.1$. 

We are interested here in the dynamics of dark energy in the gravitational field of the virialized dark matter halo. To use the system of equations of energy-momentum conservation and general relativity ones (\ref{m_cl0o})-(\ref{ee00_l2o}) we should add some 
phenomenology of virialization of matter alone or matter with dark energy. The different approaches for its description one can find
in \cite{Maor2005}. Since we concentrate our attention on the dynamics of dark energy in the bottom of gravitational well of the virialized dark matter halo, we will form the gravitational potential of virialized structure by deceleration of velocity of infall of matter to $V_m=1$ when its density contrast $1+\delta_m$ approaches to the virialization density contrast $\Delta_{vir}$. To do this we add to the left-hand side of eq. (\ref{m_cl1o}) the term for artificial viscosity or brake 
\begin{equation}
F_{vir}=\sigma_{vir}\left(\frac{\tilde{\delta}_m}{\Delta_{vir}}\right)^2, 
\label{brake}
 \end{equation}
where $\sigma_{vir}$ is the coefficient by which we set the rate of slowing. When the matter density contrast $\Delta_{vir}$ is reached at $a_{vir}$ we keep the density of matter constant, 
$\varepsilon_m=\Omega_m\bar{\varepsilon}_{cr}^{(0)}\Delta_{vir}a_{vir}^{-3}$, and amplitude of velocity $\tilde{v}_m=3aH/k^2$ by replacement of continuity and motion equations for matter (\ref{m_cl0o})-(\ref{m_cl1o}) with simple differential equations 
\begin{equation}
 \dot{\tilde{\delta}}_m-3\Delta_{vir}\frac{a^2}{a_{vir}^3}=0, \quad \dot{\tilde{v}}_m-\frac{3}{k^2}\left(H+a\frac{dH}{da}\right)=0, 
\label{m-vir}
 \end{equation}
with $\tilde{\delta}(a_{vir})=\Delta_{vir}-1$. 

\begin{figure*}
\includegraphics[width=0.48\textwidth]{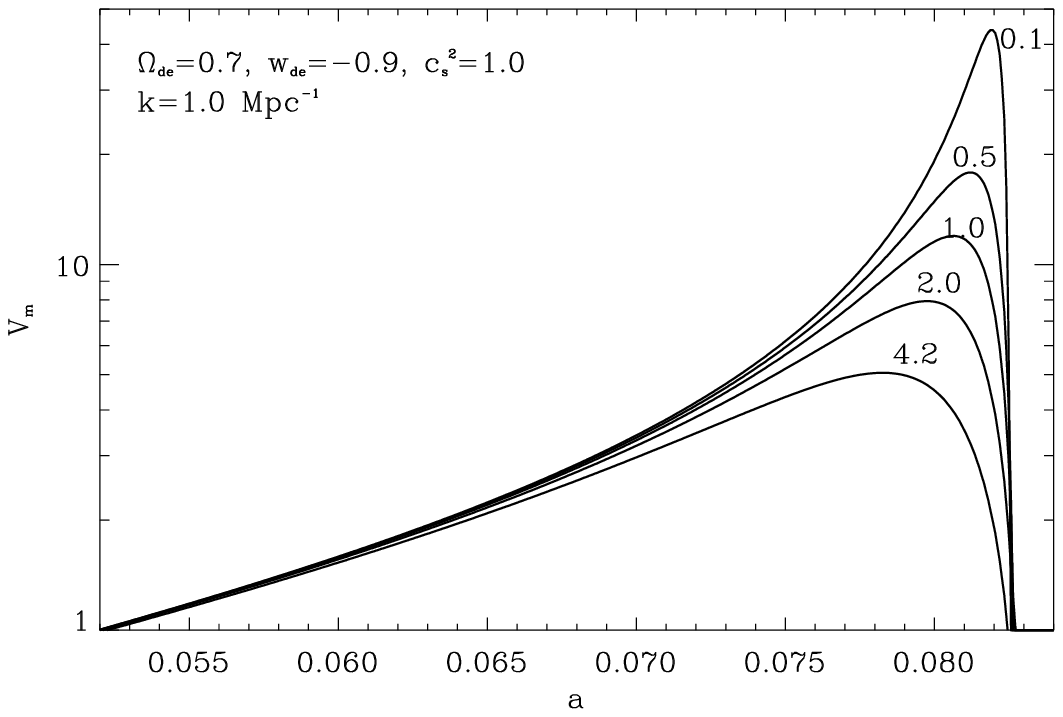} 
\includegraphics[width=0.48\textwidth]{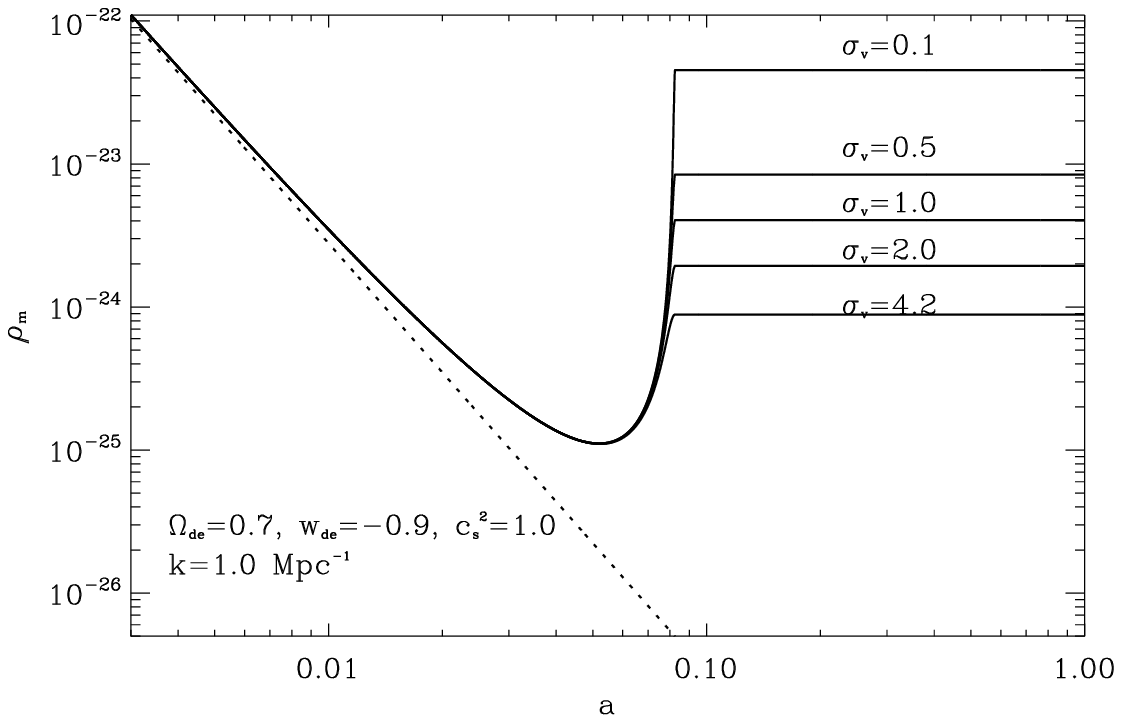} 
\caption{Virialization with different values of artificial viscosity parameter. Left: evolution of peculiar velocities of matter $V_m$ (in units of Hubble velocity) in the central part of spherical halo from turnaround point to virialization for $\sigma_{vir}=0.1,\,0.5,\,1.0,\,2.0,\,4.2$. Right: evolution of density of matter $\rho_m=\varepsilon_m/c^2$ (g/cm$^3$) in the center of spherical halo for different values of viscosity parameter. The dotted line shows the evolution of background matter density.}
\label{v-rho}
\end{figure*} 

In this section we are interested the dynamics of the halo, which are virialized at $z\sim 10$, so the good approximation is $\Delta_{vir}\approx178$ \cite{Kulinich2013,Weinberg2003}. The evolution of peculiar velocity in units of Hubble one and density of matter in the central part of spherical halo, which are computed  in the framework of phenomenology of virialization (\ref{brake})-(\ref{m-vir}) with $\sigma_{vir}=0.1,\,0.5,\,1.0,\,2.0,\,4.2$ and $\Delta_{vir}=178$, is shown in Fig. \ref{v-rho}. For lower value of $\sigma_{vir}$ we obtain the larger maximal value of peculiar velocity and higher final density in the central part of halo. The remarkable feature of such phenomenology is independence of the time interval between turnaround point and virialization, when $\varepsilon_m$ becomes constant. Comparison with linear theory (\ref{rel_cl0l})-(\ref{ee00_l2l}) shows that it practically equals the time interval from moment $a_{ta}$, when $\delta_{m}^{lin}=1.06$, to the moment, when $\delta_{m}^{lin}=1.69$,
 which practically equals $\delta_{col}$ in the Einstein-de Sitter cosmology \cite{Gunn1972,Press1974,Peebles1980,Bond1991,Bower1991,Lahav1991,Lacey1993,Eke1996,Wang1998,Cooray2002,Weller2002,Battye2003,Kulinich2003,Weinberg2003,Shaw2008,Kulinich2013}. 
It means that the latter moment corresponds to the complete collapse of the homogeneous
spherical halo of dust-like matter without virialization, $a_{col}$. So, $\Delta_{vir}$, which is defined as ratio of 
$\varepsilon_m(a_{vir})=\bar{\varepsilon}_m(a_{vir})(1+\tilde{\delta}_m(a_{vir}))$ to $\varepsilon_m(a_{col})$ is density contrast at the moment of virialization of halo. The value $\Delta_{vir}\sim178$ is reached for halo of scale $k=1$ Mpc$^{-1}$ when $\sigma_{vir}=4.2$.  
For all subsequent calculations for halo of scale $k=1$ Mpc$^{-1}$ we will keep this value. To provide the same value of $\Delta_{vir}$ for halo of other scale or other matter content,  value of $\sigma_{vir}$ must be different, one can only pick it up manually. For example, for the halo with
$k=0.5$ Mpc$^{-1}$ in the same cosmology $\sigma_{vir}=11.33$, for the halo with $k=2.0$ Mpc$^{-1}$ $\sigma_{vir}=1.43$. So, the dependence of 
value of $\sigma_{vir}$, which gives  $\Delta_{vir}\approx178$, on $k$ can be roughly approximated as
$$\sigma_{vir}\approx 4.2\left(\frac{k}{1\,{\rm Mpc}^{-1}}\right)^{-3/2}.$$

\begin{figure*}
\includegraphics[width=0.48\textwidth]{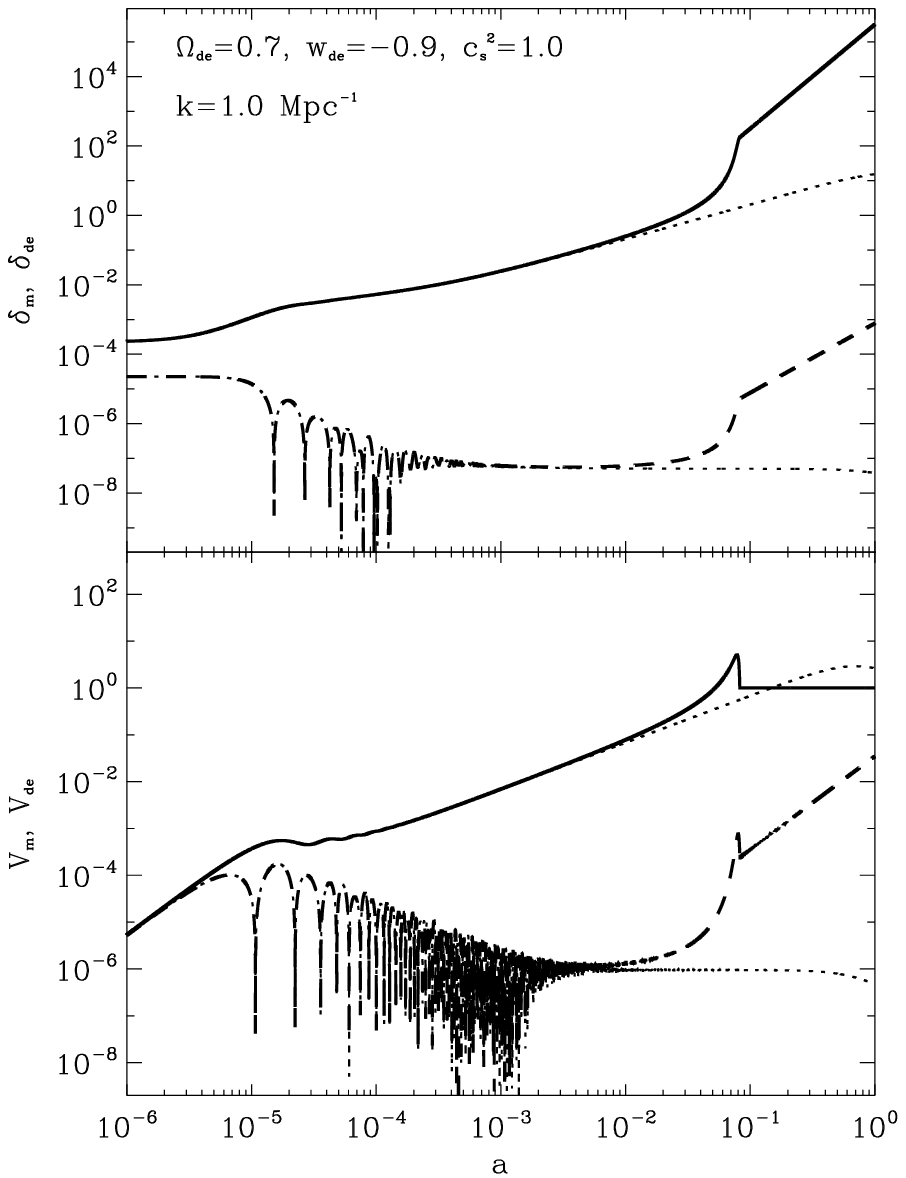}
\includegraphics[width=0.48\textwidth]{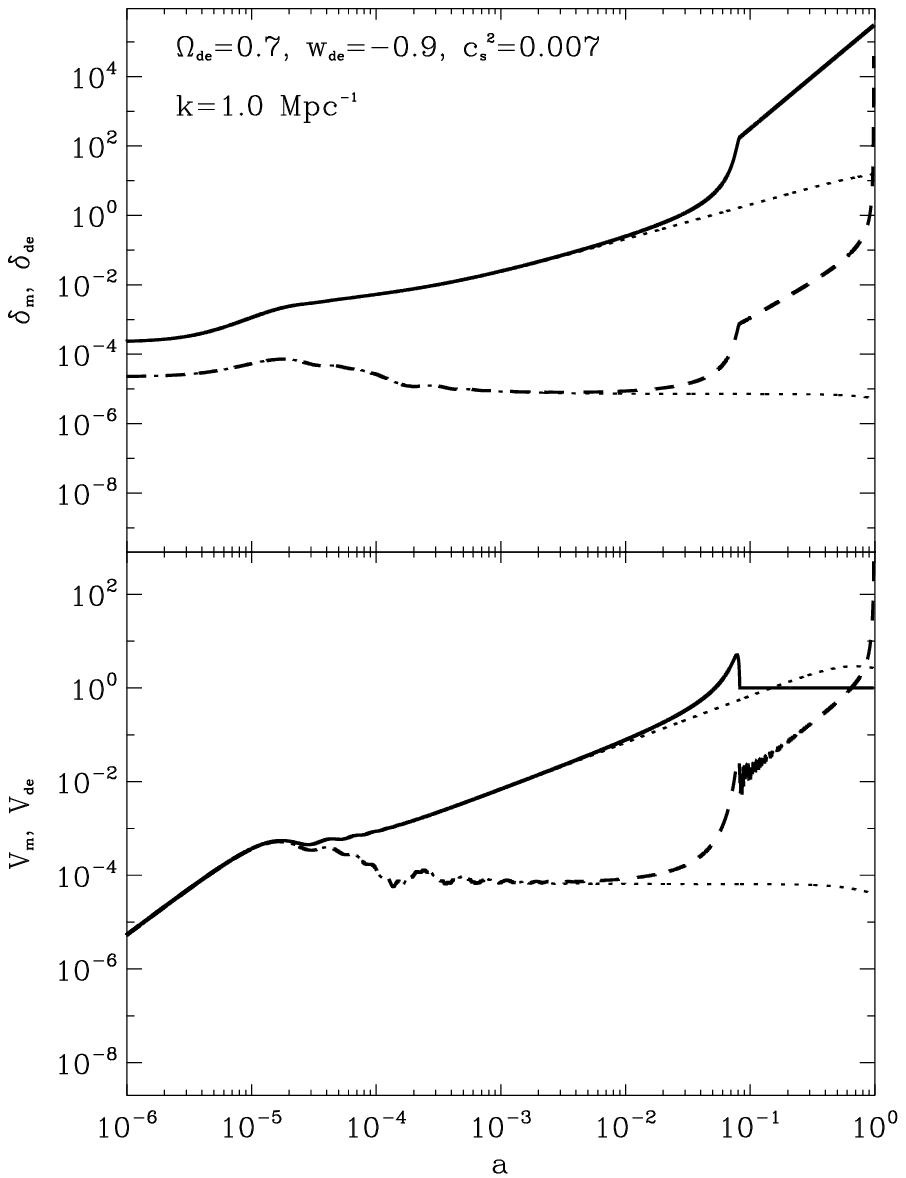}
\caption{Evolution of density perturbations of matter $\tilde{\delta}_m$ and dark energy $\tilde{\delta}_{de}$, and velocity perturbations of matter $V_m$ and dark energy $V_{de}$ in units of Hubble one at the central part of spherical halos. Solid lines corresponds to matter, dashed lines to dark energy and dotted lines show the prediction of the linear theory.}
\label{vir}
\end{figure*}

The results of integration of system of equation (\ref{m_cl0o})-(\ref{ee00_l2o}) with 
deceleration of infall of matter (\ref{brake}) which provides $\Delta_{vir}=178$ and virialized matter halo (\ref{m-vir}) are presented in the Figs. \ref{vir} for halos  with $k=1.0$ Mpc$^{-1}$ and two values of $c_s^2$: 1.0 and 0.007. The initial amplitudes we set by putting $C_k=3\cdot10^{-4}$.  

One can see that after point of turnaround of matter its contrast of density increases up ultra fast to $\approx178$ at $a_{vir}=0.083$ ($z_{vir}=11$) and after increases $\propto a^3$ to keep $\varepsilon_m=const$ (see the definition of $\delta_N(a,r)$ in sect. 2.2). For halo with $k=2.0$ Mpc$^{-1}$ it happens when $a_{vir}=0.067$ ($z_{vir}=14$) and for halo with $k=0.5$ Mpc$^{-1}$ when $a_{vir}=0.11$ ($z_{vir}=8.3$). The velocity of matter in the central part except the central point equals Hubble one, $v_m=v_H=aHr$ or $V_m=1$, that is shown in Fig. \ref{vir}. The gravitational potential $\tilde{\nu}$, which is defined mainly by the mass of matter in the halo, after virialization decreases as $\propto a^2$ in the cosmological frame (\ref{ds_sph}), or is constant in the Newtonian one. It follows from eq. (\ref{ee00_l2o}) in the Newtonian approximation. The density of matter at the central part of virialized halo $\rho_m=8.8\cdot10^{-25}$ g/cm$^3$.

The dynamics of dark energy depends on its parameters. For example, the dark energy with $\Omega_{de}=0.7$, $w=-0.9$ and $0.12<c_s\le1$ expands now in the central part of halo of matter. The amplitudes of density and velocity perturbations are small even at the current epoch: the amplitude of the density perturbation is in the range  $0.0008\le\tilde{\delta}_{de}\le0.088$ for such $c_s$ and the amplitude of the velocity perturbation in units of Hubble one is in the range $0.03\le V_{de}\le1$. They are larger for smaller $c_s$. The density of dark energy in the central part of halo 
$\rho_{de}\approx 7\cdot10^{-30}$ g/cm$^3$ and slowly decreases. So, such dark energy in the halo is practically the same as in cosmological background. The dark energy with $c_s\le0.12$ collapses into virialized halo: starts to collapse at current epoch when $c_s=0.12$ and earlier for dark energy with smaller value of $c_s$. One can see, that dark energy with $c_s=0.084$ in the halo with $k=1.0$ Mpc$^{-1}$ 
turns around at $a_{ta}\approx0.65$ ($z_{ta}\approx0.53$) and now is collapsing with $V_{de}=v_{de}/v_H\approx750$, very fast. The density of such dark energy in the center of halo equals $2.6\cdot10^{-25}$ g/cm$^3$, that is one third of the density of matter. For smaller $c_s$ the dark energy density will be larger. The dark energy in the halo with $k=2.0$ Mpc$^{-1}$ started to collapse  at $a_{ta}\approx0.94$ ($z_{ta}\approx0.06$), while in the halo with $k=0.5$ Mpc$^{-1}$ it  approaches only to the turnaround point. Such unmonotonous dependence of dynamics of dark energy on scale of halo is caused by competition of its pressure gradient and gravitation of matter of halo. The dark energy with $c_s=0$ moves together with dark matter at linear and non-linear stages. 
 
\begin{figure*}
\includegraphics[width=0.48\textwidth]{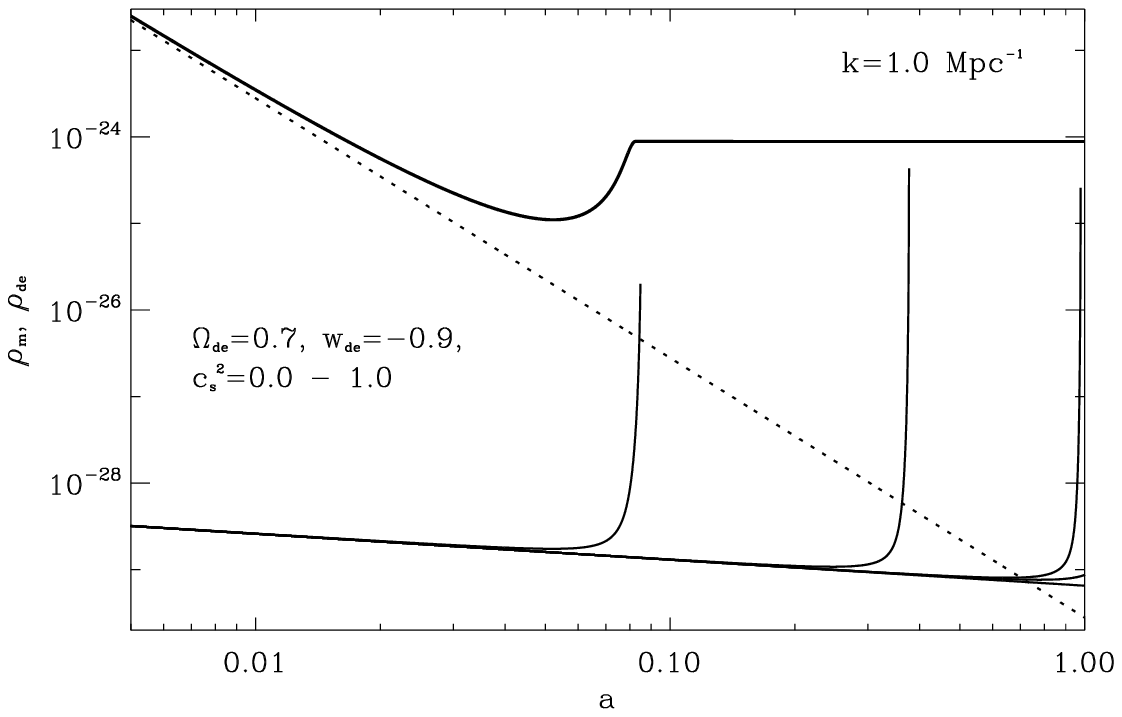} 
\includegraphics[width=0.48\textwidth]{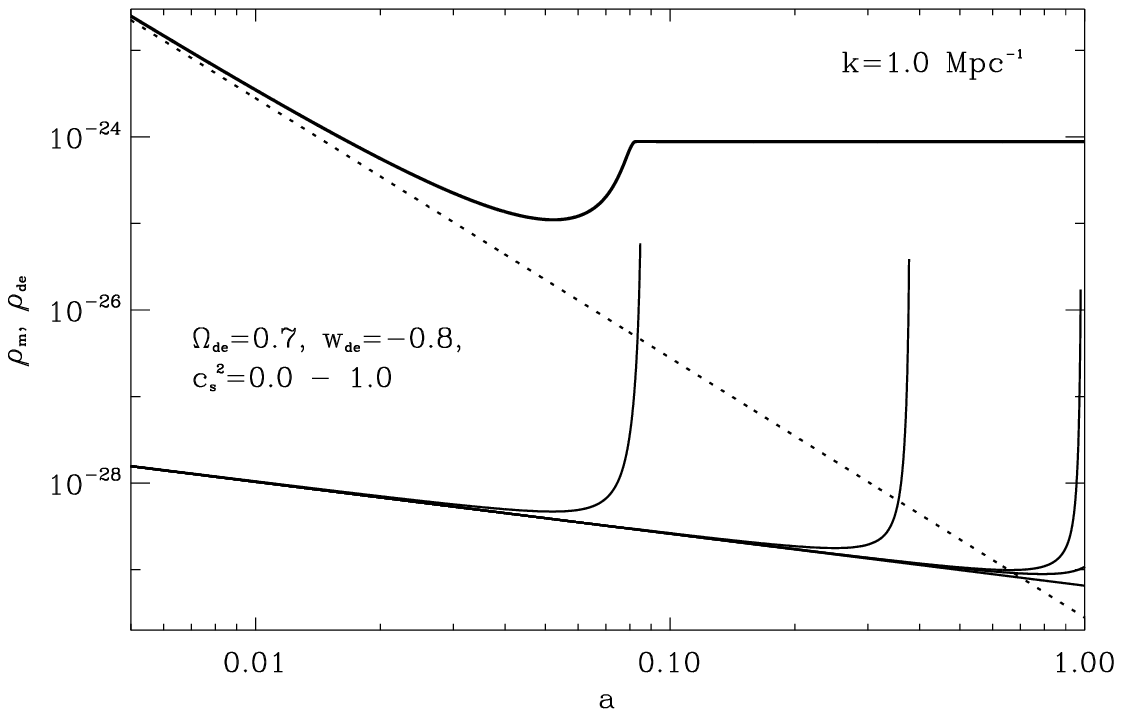} 
\caption{Evolution of density of matter $\rho_m=\varepsilon_m/c^2$ (thick solid line) and dark energy $\rho_{de}=\varepsilon_{de}/c^2$ (thin solid lines) in g/cm$^3$ at the central part of spherical halo which virialises at $z\sim10$. The density of dark energy is calculated for $c_s^2=0.0,\,0.001,\,0.007,\,
0.01,\,1.0$ (from left to right). In the left panel dark energy with $w=-0.9$, in the right one with $w=-0.8$.}
\label{rho}
\end{figure*}

In the Fig. \ref{rho} we show the evolution of matter density (see also Fig. 5.3 in \cite{Padmanabhan2002}), which reflect the features of formation of spherical halo with $k=1.0$ Mpc$^{-1}$, and evolution of density of dark energy with $c_s=0,\,0.032,\,0.084,\,0.1$ and $1.0$. It  
illustrates well that quintessential dark energy with small values of speed of sound ($c_s<0.1$) is important dynamical component of halo at late stages of their evolution: turnaround, collapse, virialization and later up to current epoch. The analysis of final stages of the evolution of such dark energy in the halo of dark matter deserves a separate paper since must include the possibility of its virialization that is out of scope of this paper.  

Presented here special characteristic values of dark energy component of halo (point of turnaround, amplitudes and so on) depend also on the other parameters of scalar field model of dark energy ($\Omega_{de}$, $w$), but that do not change the main conclusions of this paper, that follows from comparison of results presented in the left and right panels of Fig. \ref{rho}. 
 
Let's estimate the typicality of halo, which are analysed in this section, like we did in the last paragraph of previous section. The amplitude of
linear matter density perturbation with $k=1$ Mpc$^{-1}$ and $C_k=3\cdot10^{-4}$ in the same cosmology at $a=1$ is $\delta^{lin}_m=15.45$. After averaging in the top-hat sphere of halo radius $R_h=\pi/k\approx2.2$ ${\rm h^{-1}}$Mpc we obtain $\bar{\delta}^{lin}_m=4.7$. The r.m.s. density fluctuation in the scale $R_h$ is $\sigma_h\approx1.7$. Therefore, the progenitor of spherical halo analysed here is peak of height $\delta^{lin}_m/\sigma_h\approx2.8$. This density perturbation is high peak in the random Gaussian field of the initial density 
fluctuations. The galaxies and clusters have been formed from the corresponding peaks of similar heights \cite{Bardeen1986,Padmanabhan2002,Novosyadlyj2007}. 

\textit{Dynamics of dark energy in future.} We discussed the dynamics of dark energy in the halo of dark matter formation in the past and current epochs. One can see (Fig. \ref{vir}), that fluid velocity of quintessential dark energy even with $c_s\sim1$  increases, it may reach  in the future the point of turnaround ($V_{de}=1$), start to collapse and, maybe, will reach the stationary state of hydrodynamical equilibrium, discussed in \cite{Novosyadlyj2014b}. But it may happen in far future. The integration of system of equations (\ref{rel_cl0o})-(\ref{ee00_l2o}) for $a>1$ shows, that the dark energy with the same parameter as in the left panel of Fig. \ref{vir}
will reach the turnaround point at $a_{ta}\approx 7.9$ or in $\approx40$ billion years, while the dark energy with $c_s^2=0.1$ will reach 
it at $a_{ta}\approx 2.5$ or in $\approx15$ billion years.  

\section{Conclusions}

We have analysed the dynamics of scalar field dark energy in the halo of dark matter at all stages of its evolution: from linear superhorizon one, entering into horizon, evolution of scalar growing mode of perturbations in the 3-component medium up to turnaround, infall and virialization of dark matter. The last stage of matter evolution we have described phenomenologically by introducing of the artificial bulk viscosity and ``hand'' equations, which keep the density of matter constant of the given value. The evolution of dark energy at all stages is obtained by integration of the system of equations for amplitudes of perturbations in the FRW frame, which are deduced from the covariant equations of conservation laws and general relativity ones. The main conclusions are as follows:\\
i) The dynamical properties of scalar field dark energy with $w=const$ in the gravitational field of collapsing or virialized halo of dark matter strongly depend on the value of effective sound speed of dark energy. When $c_s\sim1$ the dark energy is only slightly perturbed at highly non-linear stage of evolution of dark matter halo. The amplitudes of density and velocity perturbations of such dark energy at the point of turnaround of halo are $\sim 10^{-6}$ and $\sim 10^{-4}$ accordingly. It will reach of own turnaround point in far future in few tens billion years. The models of dark energy with $c_s\le0.12$ ($\Omega_{de}=0.7$, $w=-0.9$) reach it now or in the past depending on value of $c_s$: smaller $c_s$, smaller $a_{ta}$. When $c_s=0$ then dark energy reaches the point of turnaround together with dark matter and
collapses together with it too. Such models are not excluded reliably yet \cite{Sergijenko2015,Tsizh2014}. Moreover, some authors find the evidence of their existence \cite{Lim2010,Mehrabi2015}.\\
ii) The density of dark energy with $c_s\ge0.12$ in the center of virialized halo is practically the same as in cosmological background, 
$\varepsilon_{de}/\varepsilon_m\ll1$. Such dark energy, obviously, does not affect the virialization of matter, as well as the formation of its substructure and gravitation properties of subhalos \cite{Novosyadlyj2014b}. On the contrary, the density of dark energy with $c_s\sim0$ can be large (Fig. \ref{rho}) and can essentially affect the virialization of halo, its total mass and density profile, mass function, concentration parameter and so on, that can be used for discrimination of dark energy models. Authors of \cite{Creminelli2010} discussed such possibilities recently.\\
iii) The dark energy with $w=const$ affects stronger the formation of the Great Attractor-like structures, which collapse now and are not virialized yet, the higher initial amplitude is. \\
iv) The physical and statistical properties of the most massive earliest virialized structures should discriminate the dark energy with $c_s\sim1$ and $c_s\sim0$. They are now under intensive observational and numerical simulation investigations 
\cite{Tully2014} and soon will shed more light also on the nature of dark energy, we hope.

\vskip0.5cm

\section*{Acknowledgements}
This work was supported by the projects of Ministry of Education and Science of Ukraine with state registration numbers 
0113U003059 and 0115U003279. Authors are thankful to prof. Mykola Bokalo and prof. Volodymyr Kyrylych for useful discussion on mathematical aspect of the problem and Dr. Olga Sergienko for technical help in computations. Authors also acknowledge the usage of CAMB and dverk.
\section*{Appendix~A. Non-adiabatic part of pressure perturbation in scalar field dark energy} 

 The non-adiabatic part of pressure perturbation in scalar field dark energy is caused by intrinsic entropy perturbation $\Gamma$ and is equal:
\begin{equation}
\delta p_{de}^{n-ad}\equiv p_{de}\Gamma=\delta p_{de}-w\bar{\varepsilon}_{de}\delta_{de},\label{dpnad}
\end{equation}
where $w\equiv p_{de}/\varepsilon_{de}$ and the last term is adiabatic part of pressure perturbation. The entropy perturbation can be expressed via density perturbation 
$\hat{\delta}_{de}$ and effective speed of sound $c_s^2$ in the rest frame as follows
\begin{equation}
 p_{de}\Gamma=(c_s^2-w)\bar{\varepsilon}_{de}\hat{\delta}_{de}. \label{entropy}
\end{equation}
In the rest frame ($\hat{t},\,\hat{r},\,\hat{\vartheta},\,\hat{\varphi}$) of dark energy its 4-velocity and density are as follows
\begin{equation}
\hat{u}_i=\{e^{\nu/2},\,0,\,0,\,0\}, \quad \hat{\varepsilon}_{de}(\hat{t},\,\hat{r})=\bar{\varepsilon}_{de}(\hat{t})[1+\hat{\delta}_{de}(\hat{t},\,\hat{r})], \label{u_r_f} 
\end{equation}
where $\nu$ is the same function as in (\ref{ds_sph}) but in the ``hat`` coordinates.
In the conformal Newtonian (CN) frame ($t,\,r,\,\vartheta,\,\varphi$) they are  
\begin{equation}
u_i=\{u_0,\,u_1,\,0,\,0\}, \quad \varepsilon_{de}(t,r)=\bar{\varepsilon}_{de}(t)[1+\delta_{de}(t,r)] \label{u_CN_f}
\end{equation}
where $u_0$ and $u_1$ can be expressed as in  (\ref{u-v}). In the case of small perturbations the coordinates of both frames are connected by simple transformations:
\begin{equation}
\hat{t}=t+\xi^0(t,r), \quad \hat{r}=r+\xi^1(t,r). \nonumber
\end{equation}
The density is scalar variable and transforms as 
$\bar{\varepsilon}_{de}(\hat{t})=\bar{\varepsilon}_{de}(t)+\dot{\bar{\varepsilon}}_{de}\xi^0(t,r)$. The density perturbations $\delta_{de}$ is scalar variable too, but since $\delta_{de}\ll1$ the first and next order expansion terms are second and next order of infinitesimality, so, $\hat{\delta}_{de}(\hat{t},\hat{r})\approx\hat{\delta}_{de}(t,r)$. A dot denotes the partial derivative with respect to time. Taking into account the conservation law for background density
\begin{equation}
\dot{\bar{\varepsilon}}_{de}=-3\frac{\dot{a}}{a}(1+w)\bar{\varepsilon}_{de}  \nonumber
\end{equation}
we obtain the relation
\begin{equation}
\hat{\delta}_{de}(t,r)=\delta_{de}(t,r)+3\frac{\dot{a}}{a}(1+w)\xi^0(t,r). \nonumber
\end{equation}
The unknown function $\xi^0(t,r)$ can be found from velocity transformation
\begin{equation}
u_1=\frac{\partial{\hat{t}}}{\partial{r}}\hat{u}_0+\frac{\partial{\hat{r}}}{\partial{r}}\hat{u}_1. \nonumber
\end{equation}
Taking into account (\ref{u_r_f}), (\ref{u_CN_f}) and (\ref{u-v}), we obtain
\begin{equation}
 \xi^0(t,r)=-a\int{e^{(\mu-\nu)/2}\frac{v_{de}}{\sqrt{1-v^2_{de}}}dr}.\nonumber
\end{equation}
Here we suppose also $\nu(\hat{t},\hat{r})=\nu(t,r)$, $\mu(\hat{t},\hat{r})=\mu(t,r)$ since they are small.
For inhomogeneities of galaxies and clusters of galaxies scales $\nu\ll1$, $\mu\ll1$ and $v_{de}\ll1$, so, the squared term $v^2_{de}$ in the denominator and exponent can be omitted, so 
\begin{equation}
 \hat{\delta}_{de}(t,r)=\delta_{de}(t,r)-3\dot{a}(1+w)\int{v_{de}dr}. \label{de_transf}
\end{equation}
Gathering all together (\ref{dpnad}), (\ref{entropy}) and (\ref{de_transf}) we find
\begin{equation}
\delta p_{de}=\bar{\varepsilon}_{de}\left[c_s^2\delta_{de}-3\dot{a}(1+w)(c_s^2-w)\int{v_{de}dr}\right].  \label{delta_p}
\end{equation}
Other approaches and details of deducing of the non-adiabatic part of pressure perturbation of scalar field can be found in \cite{Hu1998,Hu2004,Gordon2004,Unnikrishnan2008}. The contribution of non-adiabatic part of pressure perturbation (last term in (\ref{delta_p}), fifth term in (\ref{de_cl0}) and fourth term in (\ref{de_cl0o}) is important at superhorizon linear stage and practically disappears when perturbation enters into horizon. That is why we do not generalized this term for the non-linear stage.  

At last we would like to note here, that some researchers use the term ''effective sound speed of dark energy``  as speed of propagation 
of perturbation in any frame (see, for example, \cite{Abramo2009}). It is related with our $c^2_s$ by simple equation
$c^2_{eff}=c_s^2-3\dot{a}(1+w)(c_s^2-w)\tilde{v}_{de}/\tilde{\delta}_{de}.$ Therefore, $c^2_{eff}$ is variable for constant $c^2_s$.

\section*{Appendix~B. Linearized system of differential equations for evolution of perturbations in three-component medium}
 
The system of ordinary differential equations which describes the evolution of Fourier amplitudes of cosmological linear perturbations of metric, densities and velocities of three-component zero-shear medium in the conformal-Newtonian gauge is as follows
\begin{eqnarray}
&&\dot{\tilde{\delta}}_{\rm r}-2\dot{\tilde{\nu}}-\frac{4k^2}{3a^2H}\tilde{v}_{\rm r}=0,\quad
\dot{\tilde{v}}_{\rm r}+\frac{\tilde{\nu}}{2a^2H}+\frac{\tilde{\delta}_{\rm r}}{4a^2H}=0,\label{rel_cl0l}\\
&&\dot{\tilde{\delta}}_{m}-\frac{3}{2}\dot{\tilde{\nu}}-\frac{k^2}{a^2H}\tilde{v}_{m}=0,\quad
\dot{\tilde{v}}_{m}+\frac{\tilde{v}_{m}}{a}+\frac{\tilde{\nu}}{2a^2H}=0,\label{m_cl0l}\\
&&\dot{\tilde{\delta}}_{de}+\frac{3}{a}(c_s^2-w)\tilde{\delta}_{de}-(1+w)\left[\frac{k^2}{a^2H}\tilde{v}_{de}
+9H(c_s^2-w)\tilde{v}_{de}+\frac{3}{2}\dot{\tilde{\nu}}\right]=0, \label{de_cl0l}\\
&&\dot{\tilde{v}}_{de}+(1-3c_s^2)\frac{\tilde{v}_{de}}{a}+\frac{c_s^2\tilde{\delta}_{de}}{a^2H(1+w)}+\frac{\tilde{\nu}}{2a^2H}=0, \label{de_cl1l}\\
&&\dot{\tilde{\nu}}+\left(1+\frac{k^2}{3a^2H^2}\right)\frac{\tilde{\nu}}{a}=
-\frac{H^2_0}{H^2}\left(\Omega_ma^{-3}\tilde{\delta}_m+
\Omega_{\rm r}a^{-4}\tilde{\delta}_{\rm r}+\Omega_{de}a^{-3(1+w)}\tilde{\delta}_{de}\right). \label{ee00_l2l} 
\end{eqnarray}
The system has well known analytical solutions for two special cases -- radiation-dominated epoch ($\Omega_{\rm r}=1$, $\Omega_m=\Omega_{de}=0$) and matter-dominated one ($\Omega_{m}=1$, $\Omega_{\rm r}=\Omega_{de}=0$). For two or three component case it can be solved numerically
for given initial conditions, for which we have designed the FORTRAN routine dedmhalo-l.f.  

\begin{figure*}[h!]
\begin{centering}
\includegraphics[width=0.6\textwidth]{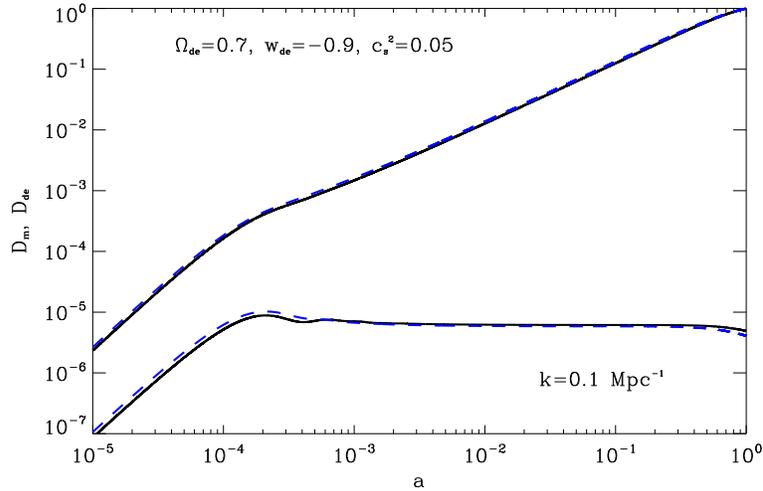}
\caption{Evolution of density perturbations of matter and dark energy (transformed to synchronous gauge) obtained by integration of the system of equations (\ref{rel_cl0l})-(\ref{ee00_l2l}) by our code dedmhalo-l.f (solid black lines), and the same obtained by integration of Boltzman-Eistein code CAMB in synchronous gauge (dasched blue lines).}
\end{centering}
\label{camb}
\end{figure*}
The results of numerical integration of this system of equations with initial conditions (\ref{dr_ini})-(\ref{de_ini}) are presented in the Figs. (\ref{camb}) and (\ref{vir}) by dotted lines. In Fig. \ref{camb} we present the evolution of density perturbations of dark matter and dark energy obtained by integration of system of equation (\ref{rel_cl0l})-(\ref{ee00_l2l}) by our code dedmhalo-l.f 
and transformed to synchronous gauge according to known relation $D_N=\tilde{\delta}_N-3(1+w_N)v_N$ \cite{Bardeen1980}. The evolution of $D_m$ and $D_{de}$ obtained by integration of system of Boltzman-Eistein equations in synchronous gauge by code CAMB \cite{camb} is also presented there. This illustrates well agreement of results obtained in the different approaches and by different codes.

\end{document}